\documentclass[prd,twocolumn,amsmath,amsfonts,amssymb,showpacs,nofootinbib,floatfix]{revtex4} 
\usepackage{epsfig,psfrag,graphicx,bm,color,xspace,enumitem} 
\usepackage{lipsum}
\usepackage{footmisc}
\usepackage{graphicx,color}
\usepackage[abs]{overpic}

\newcommand{\sss}{\scriptscriptstyle} 
\newcommand{\formcalc}{\textsf{FormCalc}\xspace} 
 
\newcommand{\ds}{\textsf{DarkSUSY}\xspace} 
\newcommand{\micro}{\textsf{micrOMEGAs}\xspace} 
\newcommand{\mg}{\textsf{MadGraph}\xspace} 
\newcommand{\me}{\textsf{MadEvent}\xspace} 
\newcommand{\pythia}{\textsf{Pythia}\xspace} 
 
\newcommand{\helas}{\textsf{HELAS}\xspace} 
\newcommand{\pgs}{\textsf{PGS}\xspace} 
\newcommand{\gfitter}{\textsf{Gfitter}\xspace}

\newcommand{\beq}{\begin{equation}} 
\newcommand{\eeq}{\end{equation}} 
\newcommand{\bea}{\begin{eqnarray}} 
\newcommand{\eea}{\end{eqnarray}} 
\newcommand{\nn}{\nonumber} 
\newcommand{\refeq}[1]{(\ref{#1})} 

\def\ET{{\displaystyle{\not}E_T}}

\begin{document} 

\author{Michael Gustafsson} 
\email{mgustafs@ulb.ac.be} 
\affiliation{Service de Physique Th\'eorique, Universit\'e Libre de Bruxelles, CP225, Bld du Triomphe, 1050 Brussels, Belgium}  

\author{Sara Rydbeck} 
\email{sara.rydbeck@desy.de} 
\affiliation{Deutsches Elektronen-Synchrotron DESY, Notkestra\ss e 85, D-22607 Hamburg, Germany}

\author{Laura Lopez-Honorez}
\email{llopezho@mpi-hd.mpg.de}
\affiliation{Max-Planck-Institut fuer Kernphysik, Saupfercheckweg 1,  69117 Heidelberg, Germany}

\author{Erik Lundstr\"om}
\affiliation{The Oskar Klein Centre, 
AlbaNova University Center, SE - 106 91 Stockholm, Sweden}
\thanks{Former academic affiliation.}

\date{\today} 

\title{Status of the inert doublet model and the role of multileptons at the LHC} 
\vspace{-4.5ex}
{\normalsize \raggedright
DESY 11--261\\
ULB-TH/12-08\\[10ex]
}

\begin{abstract} 
A possible feature of the inert doublet model (IDM) is to provide a dark matter candidate together with an alteration of both direct and indirect collider constraints that allow for a heavy Higgs boson. We study the IDM in light of recent results from Higgs searches at the Large Hadron Collider (LHC) in combination with dark matter direct detection limits from the XENON experiment. We ask under what conditions the IDM can still accommodate a heavy Higgs boson. We find that IDM scenarios with a Higgs boson in the mass range 160-600 GeV are ruled out only when all experimental constraints are combined. For models explaining only a fraction of the dark matter the limits are weakened, and IDMs with a heavy Higgs are allowed. We discuss the prospects for future detection of such IDM scenarios in the four-lepton plus missing energy channel at the LHC.  This signal can  show up in the first year of running at $\sqrt{s}= 14$ TeV, and we present detector-level studies for a few benchmark models.
\end{abstract} 

\maketitle

\section{Introduction} 
The era for studying particle physics with the LHC at CERN is ongoing. Since 2010, the experiments have been collecting data from proton-proton collisions at a center-of-mass energy of 7 TeV. This has already enabled the exploration of new regimes of the current standard model (SM) as well as physics beyond the SM. One of the aims is to establish or exclude the presence of a SM Higgs boson\footnote{The term Higgs boson will throughout the text be used for the physical scalar particle emerging from the electroweak symmetry breaking in the SM by the Brout-Englert-Higgs mechanism [F. Englert and R. Brout, Phys. Rev. Lett. 13, 321 (1964); P.W. Higgs, Phys. Rev. 145, 1156 (1966)].}. The latest public Higgs search results were presented by the ATLAS and CMS collaborations in March 2012~\cite{atlas-combined-update,cms-combined-update}. These analyses exclude a SM Higgs boson in the range $127-600$ GeV to the 95\% confidence level (CL). It is however important to keep in mind that new particles can both contribute to the Higgs decay width and alter its production cross section. The exclusion limits on this range of Higgs boson masses might thus not be valid for a Higgs boson that is SM-like in many respects, but which also couples to states beyond the SM. This is of particular relevance to the present paper. Let us point out that while both the ATLAS and CMS experiments have started to see potential evidence for a particle signal at $\sim125$~GeV, the significance is not yet enough to claim discovery and establish this to be caused by the Higgs particle itself. Moreover, there have been other, perhaps interesting, excesses in the Higgs searches; {\it e.g.}\ at the $2\sigma$ level for a $\sim320$~GeV particle mass in CMS \cite{Chatrchyan:2012dg, Chatrchyan:2012tx}, which was then not confirmed by the latest preliminary results from the ATLAS experiment \cite{ATLAS:2012ac}.

Because of the nature of hadron colliders, the LHC has obvious advantages in probing beyond the SM scenarios that incorporate strong quantum chromodynamic (QCD) interactions, such as minimal low-energy supersymmetry models. So far the LHC searches have found no evidence for strongly interacting beyond the SM particles \cite{susyATLASCMS}. Notoriously, scenarios without direct SM QCD interactions are expected to give lower signals -- although many exceptions, such as resonances ({\it e.g.}\ \cite{Salvioni:2009jp}) or composite state effects ({\it e.g.}\ \cite{Belyaev:2008yj,Chivukula:2011ue}), may appear.  From an empirical point of view, there is {\it a priori} no need for new QCD interacting sectors. Indeed, two of the major questions in particle physics and cosmology -- the fine-tuning problem in the SM Higgs sector (commonly known as the ``LEP paradox" or the hierarchy problem \cite{Barbieri:2000gf}), and the dark matter (DM) problem with a thermally produced weakly interacting massive particle (WIMP) as one of the longstanding candidate solutions \cite{Jungman:hep-ph9506380,Bergstrom:2000pn,Bertone:2004pz,Cirelli:2012tf,Bergstrom:2012fi} -- are not directly connected to  QCD properties. 

\smallskip
Given the latter point of view, we study the inert doublet model (IDM); a minimal extension of the SM which contains one additional electroweak scalar doublet and has the potential both to alleviate the mentioned fine-tuning in the SM and to provide a DM candidate.
The IDM appeared already in the 1970s in \cite{Deshpande:1977rw}, but received more attention after Barbieri {\it et.al.} \cite{Barbieri:2006dq} (see also \cite{Ma:2006km}) showed that the model could provide both a DM candidate by an imposed $Z_2$ symmetry and allow for SM-like Higgs masses up to 600 GeV without contradicting electroweak precision data. These authors pointed out how raising the Higgs mass could alleviate the problem posed by the LEP paradox \cite{Barbieri:2006dq} and thus eliminate the fine-tuning in the SM up to an energy scale of a few TeV (see however \cite{Casas:2006bd}).  
Regarding its DM candidate \cite{LopezHonorez:2006gr, Gustafsson:2007pc, Dolle:2009fn}, many signatures have been studied and range from potentially striking gamma-ray lines \cite{Gustafsson:2007pc}, cosmic-ray and neutrino fluxes \cite{Agrawal:2008xz, Nezri:2009jd} to direct detection signals \cite{Majumdar:2006nt,LopezHonorez:2006gr,Arina:2009um}. 
The lack of conclusive beyond the SM signals in these channels and in the data from collider experiments \cite{Barbieri:2006dq,Cao:2007rm,Gustafsson:2007pc,Lundstrom:2008ai} has so far only partially constrained the IDM parameter space. 

\medskip 
We devote the first part of this paper to the question of whether a large Higgs boson mass ($\gtrsim160$ GeV) can still be accommodated within the IDM. Indeed, fairly large inert particles-Higgs couplings are needed in order to elude the SM Higgs searches at the LHC. The same couplings are, however, severely constrained by DM direct detection searches at XENON-100. 

The need for large scalar couplings leads us to the second part of the paper, where we study a new potential discovery channel for the IDM in the form of multilepton events via 
 heavy Higgs production at the LHC.

Even if the picture of the Higgs being SM-like will become clearer as the LHC continues to run during 2012, the possibility and the nature of a modified Higgs sector might remain an open question. After 2012, the LHC will have a long shutdown in preparation for start-up in late 2014 at the design center-of-mass energy of 14 TeV. Once the design luminosity is reached, the experiment will accumulate $\sim100$~fb$^{-1}$ per year, allowing probing of the Higgs sector and beyond SM physics in more detail.

The prospects for detecting IDM signatures in the upcoming LHC data at 14 TeV has already been partly explored. In \cite{Barbieri:2006dq,Cao:2007rm} the authors studied how inert particles affect SM Higgs searches, by the opening of additional decay channels, as well as the discovery potential in the dilepton and missing energy channel. A more comprehensive study of this dilepton channel was done in \cite{Dolle:2009ft}, followed by a trilepton study \cite{Miao:2010rg}. None of these studies explore the possibility to detect the inert doublet model in the almost background-free multilepton ($\geq 4$ leptons) plus $\ET$ channel. Here we argue that it is natural to study the tetralepton channel in addition to the dilepton and trilepton channels.  This has actually been done for many other popular models, {\it e.g.}\ in supersymmetry \cite{Baer:1992kd,Baer:1994fx,Moortgat:2001pp,Bisset:2007mi,Gentile:2009zz} and extra dimension \cite{Cheng:2002ab,Kazana:2007zz} models. 

The inert doublet contains four new particle states. The more massive states may be pair produced in proton collisions and subsequently cascade decay (in one or two steps) down to the lightest inert particle state, which remains stable due to the conserved $\mathbb Z_2$ parity. In each decay step, an electroweak gauge boson is produced and can decay into one or two charged leptons. If the lightest stable inert particle is electrically neutral, it will contribute to the missing transverse energy ($\ET$), and up to six charged leptons can be directly produced from the $W^\pm$ and $Z$ boson that participated in the cascade decay. We show that the $(\geq4l+\ET)$ channel is an interesting test of the IDM and can provide an early discovery channel of the IDM when the LHC runs at 14 TeV. 

\medskip	
In Sec.~\ref{sec:1} and \ref{sec:2} we set up the IDM framework and the theoretical, experimental and observational constraints that will be imposed on the model. In Sec.~\ref{sec:3} we answer our first question, namely under what conditions a heavy SM-like Higgs can survive the recent and complementary constraints from the LHC and XENON. In Sec.~\ref{sec:4} we turn to our second aim, to discuss the multilepton signal at the LHC in such scenarios. We perform detailed event simulations for a set of IDM benchmark models and the SM background and describe our analysis tools in Sec.~\ref{sec:5}.  Our results and discovery prospects for IDM in the tetralepton+$\ET$ channel are presented in Sec.~\ref{sec:6}, and in Sec.~\ref{sec:7} we summarize and conclude.

\section{The inert doublet model}\label{sec:1} 
The IDM consists of the SM, including the standard Higgs doublet $H_1$, and an additional Lorentz scalar in the form of an SU(2)$_L$ doublet $H_2$. An extra unbroken $\mathbb Z_2$ symmetry is introduced, under
which $H_2$ is taken to be odd ($H_2 \rightarrow - H_2$) while $H_1$ and all other SM fields are even. This $\mathbb Z_2$ symmetry protects against the introduction of new flavor changing neutral currents and guarantees the absence of direct Yukawa couplings between the inert states and the SM fermions (hence the name \emph{inert} doublet model). The symmetry also renders the lightest particle state of $H_2$ stable. If neutral, the latter can provide a good DM candidate. The new kinetic gauge term takes the usual form, $D^\mu H_2 D_\mu H_2$, and the most general renormalizable CP conserving potential for the IDM scalar sector is
\bea
\label{eq:potential}
V&=& \mu_1^2 \vert H_1\vert^2 + \mu_2^2 \vert H_2\vert^2+ \lambda_1 \vert H_1\vert^4 + \lambda_2 \vert H_2\vert^4
\nn\\ \noalign{\vskip 2mm}
  && \mbox{ }\hspace{-1.0cm}  +\; \lambda_3 \vert H_1\vert^2 \vert H_2 \vert^2 + \lambda_4 \vert H_1^\dagger H_2\vert^2 + {\lambda_5} Re[(H_1^\dagger H_2)^2],
\eea

\noindent
where $\mu_i^2$ and $\lambda_i$ are real parameters. 

Four new physical particle states are obtained in this model: two charged states, $H^{\pm}$, and two neutral states, $H^0$ and $A^0$.  After standard electroweak symmetry breaking, the masses of the scalar particles (including the SM-like Higgs mass $m_h$) are given by:
\bea \label{eq:masses}
  m_{H^0}^2   &=& \mu_2^2 + (\lambda_3 + \lambda_4 + \lambda_5) v^2 \;\equiv\; \mu_2^2 + \lambda_{H^0} v^2,\nn\\
  m_{A^0}^2   &=& \mu_2^2 + (\lambda_3 + \lambda_4 - \lambda_5) v^2  \;\equiv\; \mu_2^2 + \lambda_{A^0} v^2,\nn\\
  m_{H^\pm}^2 &=& \mu_2^2 +  \lambda_3 v^2,\nn\\
     m_h^2   &=& -2\mu_1^2=4\lambda_1v^2,
\eea
where $v\approx177$ GeV is the vacuum expectation value of the Higgs field $H_1$. In the following, we choose $H^0$ to be the lightest inert particle, and hence the potential DM candidate. Notice that the roles of $A^0$ and $H^0$ are equivalent in the IDM and our conclusions would remain unchanged if we had chosen $A^0$ to be the DM candidate. A convenient set of parameters to describe the full scalar sector are the four scalar masses  $\{m_{H^0},m_{A^0},m_{H^\pm},m_h\}$, the self-coupling $\lambda_2$ and $\lambda_{H^0} \equiv \lambda_3 + \lambda_4 + \lambda_5$.

\section{Constraints on IDM}\label{sec:2}
There are several theoretical, experimental and observational constraints on the model that have to be considered. For all the models in this study we consistently impose: 
\smallskip
\begin{enumerate}[topsep=4pt, partopsep=2pt,itemsep=4pt,parsep=2pt,leftmargin=1.9em]
  \item[$\bullet$] the requirements for vacuum stability \cite{Ginzburg:2010wa, Gustafsson:2011wg},
  \item[$\bullet$] that calculations should be within the perturbative regime (with $\lambda_i < 4\pi$) \cite{Barbieri:2006dq,Gustafsson:2011wg}\footnote{The constraint in Eq.\;17 of reference \cite{Barbieri:2006dq}, that poses a sufficient condition not to affect their naturalness arguments for the IDM, is not included. Applying it does not change our conclusions, although it would reject the models in our scans which have $m_{H^0} \gtrsim 120$~GeV and correct relic density $\Omega_{H^0} \approx \Omega_\mathrm{CDM}$.},
  \item[$\bullet$] unitarity constraints (the absolute value of the eigenvalues of the $S$ matrix are required to be $\leq$ 1/2 for scalar-to-scalar scatterings, including the longitudinal parts of the gauge bosons)
   \cite{Lee:1977yc,Lee:1977eg,Arhrib:2000is,Akeroyd:2000wc,Ginzburg:2005dt,Eriksson:2009ws}\footnote{See also \cite{Gorczyca:2011rs}, where the authors studied the constraints from unitarity on the IDM.},
  \item[$\bullet$] consistency with electroweak precision tests (EWPT) (99\% CL) \cite{Baak:2011ze},
  \item[$\bullet$] consistency with particle collider data from LEP ($\sim$95\% CL) \cite{Barbieri:2006dq,Cao:2007rm,Pierce:2007ut,Lundstrom:2008ai},  
  \item[$\bullet$] a relic abundance of $H^0$ in agreement with the WMAP measured $\Omega_\mathrm{CDM}h^2=0.1109\pm0.017$ ($3\sigma$) \cite{Larson:2010gs},
  \item[$\bullet$] consistency with direct DM searches by XENON (90\% CL) \cite{Angle:2011th,Aprile:2011},   
  \item[$\bullet$] consistency with indirect DM searches. We include the 95\% CL gamma-ray constraints by Fermi-LAT (assuming Navarro-Frenk-White profiles) \cite{Abdo:2010nc,Ackermann:2011wa,Ackermann:2012qk}.  No other indirect detection probes are considered here as these either give significantly weaker limits or are associated with too large astrophysical uncertainties.  
\end{enumerate} 
\smallskip
IDMs with large Higgs masses can potentially alleviate the fine-tuning present in the SM and thus address the LEP paradox \cite{Barbieri:2006dq}. While we choose not to impose any explicit naturalness constraints here, we will extensively comment on this in Appendix \ref{sec:naturalness}. 

For a review of many of the constraints on IDM we refer to \cite{Gustafsson:2011wg}. We have implemented the above list of constraints, as described in the given references, into our own computer code. We stress the importance of combining all these bounds since, as we will see, their complementarity becomes a powerful tool in constraining the IDM.

\smallskip
We will present results of random scans over the whole viable IDM parameter space that is of interest for our study (from a few GeV to hundreds of GeV). More precisely, the free parameters were taken to be the three masses of the inert scalars, the Higgs mass and the coupling $\lambda_L$. We scanned over the ranges:
\bea
  5 \mbox{ GeV}   \leq &m_{H^0}& \leq 170 \mbox{ GeV}, \nn\\
 m_{H^0} \leq &m_{A^0}& \leq 800 \mbox{ GeV} ,\nn\\
 \mbox{max}(m_{H^0}, 70  \mbox{ GeV} )\leq &m_{H^{\pm}}& \leq 800 \mbox{ GeV} ,\nn\\
100 \mbox{ GeV}  \leq &m_{h}& \leq 900 \mbox{ GeV} \nn\\
 10^{-5}\leq  &|\lambda_L|& \leq 4\pi.\nn
\eea
Once these parameters were chosen randomly, the value of $\lambda_2$ was fixed to its minimal value satisfying the constraints from vacuum stability. The resulting IDMs were confronted with the constraints listed in this section and only models passing the full set of constraints were considered as viable.  

A random scan is always incomplete in covering all possible models. By a combination of random scans, simple Markov chain Monte Carlo  searches (following~\cite{Gondolo:2004sc}) and physical insight into where models could be expected to be found, we believe that we have been able to cover all relevant parts of the parameter space for our results with more than 100\,000 models. For example, earlier studies  \cite{Hambye:2009pw,LopezHonorez:2010tb}  have already shown that expanding the scan to larger $H^0$ masses is not relevant if $H^0$ should constitute a WIMP DM candidate. This is at least true for $H_0$ masses below $500$ GeV, and higher masses are not relevant for the current LHC searches. It is worth noting that this part of the IDM gives well isolated regions in all our presented quantities. In practice, no viable IDMs were found with $m_h\gtrsim 700$ GeV, $m_{H^0}\gtrsim 150$ GeV, $m_{A^0}\lesssim 50$ GeV or $|\lambda_{H^0}|\gtrsim 7$. 

The DM relic density calculations have been performed by \ds \cite{Gondolo:2004sc} interfaced with \formcalc \cite{Hahn:1998yk}. This code was originally developed in \cite{Gustafsson:2007pc}, but has now been updated to also include three-body final states (as in \cite{Honorez:2010re}). Also, an upgrade of \micro~\cite{Belanger:2008sj} including annihilation into three-body final states \cite{Honorez:2010re} has been used for the scans.

\section{IDM in light of XENON and the LHC Higgs search}\label{sec:3}
Dark matter direct detection and the LHC's SM Higgs searches are known to be complementary in constraining Higgs portal DM models \cite{Farina:2011bh,Mambrini:2011ik,Raidal:2011xk,Baek:2011aa,Djouadi:2011aa,He:2011gc,Lebedev:2011iq,LopezHonorez:2012kv,Djouadi:2012zc}. Direct detection experiments pose upper limits on the DM coupling to the Higgs. This in turn restricts the Higgs decay rate into the invisible DM states, which makes it more difficult for such models to escape the bounds coming from LHC's Higgs particle searches.

The constraints on singlet scalar DM from combining XENON-100 and the LHC SM Higgs searches were {\it e.g.}\ studied in \cite{Mambrini:2011ik} for a wide range of Higgs masses. Let us emphasize that the latter analysis did not assume any explicit mechanism for evading EWPT constraints, which would otherwise constrain the SM Higgs mass to be below roughly 160 GeV. By contrast, the IDM provides such a mechanism and can easily accommodate Higgs masses up to at least 600 GeV while still being in agreement with EWPT. Another difference to the singlet scalar DM model is that the IDM's ``dark" sector is composed of more than one particle state. The additional states potentially provide new contributions to the decay width of the SM-like Higgs boson, along with additional processes relevant for the determination of the DM relic density.

\subsection{Constraints from direct detection DM searches}\label{sec:dd}
\begin{figure}[t]
\includegraphics[width=1.05 \columnwidth]{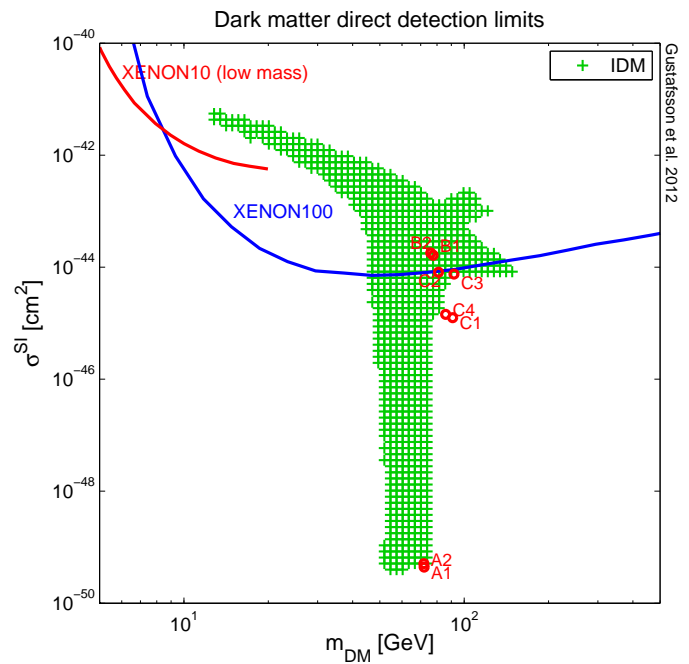}
\caption{Direct detection signal for IDMs in agreement with WMAP data. 
The crosses represent IDM models that pass all our imposed constraints from Sec.~\ref{sec:2} (not including LHC constraints). The upper 95 \% CL bounds from XENON-100 \cite{Aprile:2011} and XENON-10 \cite{Angle:2011th} are shown by the solid lines. 
The points labeled A1-C4 show the scattering cross section for our benchmark models. As explained in Sec.~\ref{sec:5}, the B models can only pass the constraints after taking systematic uncertainties into account, and the C models account only for a fraction of the DM density. 
}\label{fig:dd}
\end{figure}
Figure \ref{fig:dd} shows how XENON \cite{Aprile:2011,Angle:2011th} constrains the IDM models that have a relic density in agreement with WMAP. These constraints assume a local $H^0$ density of $\rho_0=0.3$ GeV/cm$^{-3}$ and a standard Maxwellian velocity distribution. The spin-independent cross section for IDMs is calculated as in
\cite{Barbieri:2006dq}:
\beq
\sigma^{\text{SI}}_{H^0\text{-}p}=\frac{m_n^4\lambda_{H^0}^2f^2}{
{4}\pi(m_n+m_{H^0})^2m_h^4}\label{eq:SSI},
\eeq
where the form factor is taken to be $f=0.3$ \cite{Ellis:2000ds,LopezHonorez:2006gr,Mambrini:2011ik}, and $m_n$ is the target nucleon mass.  The loop induced contribution estimated in \cite{Barbieri:2006dq} is also included, but it is very small. 

This leaves a viable mass range roughly between 45 and 80 GeV for the DM candidate $H^0$. The range can be extended up to $\sim$~150 GeV with a few models marginally surviving the current XENON-100 bound \cite{LopezHonorez:2010tb}. The low mass region below $\lesssim 10$ GeV is excluded both by XENON-10 \cite{Angle:2011th} and by Fermi-LAT gamma-ray constraints \cite{Ackermann:2011wa,Andreas:2010dz}.\footnote{Concerning a low mass WIMP, there is a debate as to what extent the exclusion limits from direct detection results are reliable (see {\it e.g.}\ \cite{Collar:2010-2011}). In order to be conservative, we could therefore choose not to include the XENON-10 upper bounds. At the same time, we note that the WIMP signal constraints from the Fermi-LAT data on gamma-rays from {\it e.g.}\ dwarf galaxies \cite{Ackermann:2011wa,Andreas:2010dz} also exclude this low $H^0$ mass region of the IDM. We therefore take the viewpoint that a light $H^0$ below 10~GeV is not a viable WIMP candidate within the current standard scenario \cite{Gustafsson:2011wg}.} We will, however, include low $H^0$ masses in parts of the following discussion for illustrative purposes, although they are excluded once we impose all our constraints.

A viable large $H^0$ mass region above $\sim$500 GeV also exists \cite{Hambye:2009pw}, but is not of interest for the present study. Such heavy IDM states would for kinematical reasons never alter the width of the Higgs boson (with a mass below 1~TeV) and therefore the LHC constraints apply exactly as in the SM. Such heavy IDM states will also be very difficult to probe directly at the LHC.  On top of that, in order to get the correct relic density and  to comply with EWPT, only small Higgs masses can be considered~\cite{Hambye:2009pw}.

\subsection{Constraints from Higgs boson searches}

\begin{figure*}
\centering
\begin{tabular}{cc}
\includegraphics[width=1.0 \columnwidth]{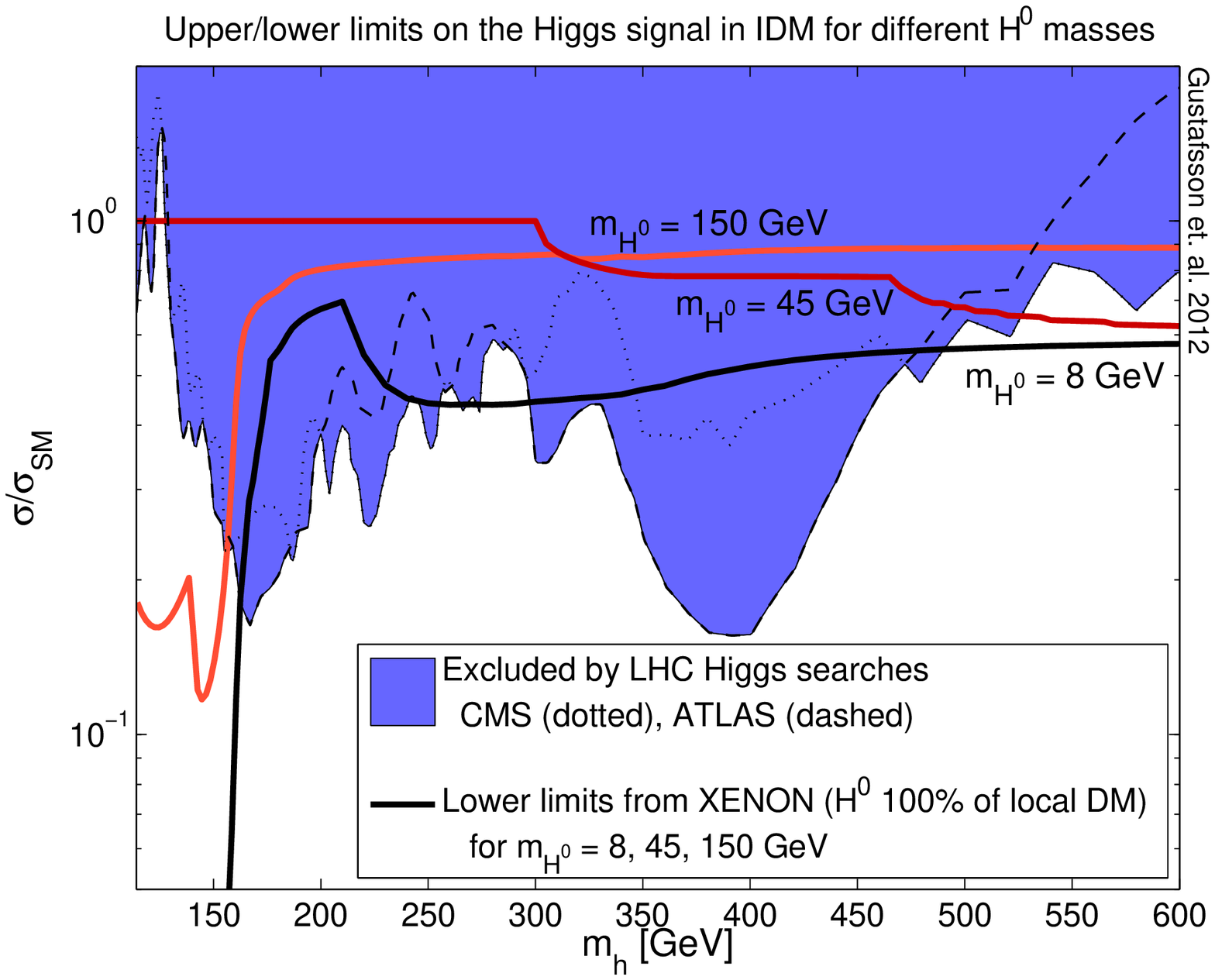}
\includegraphics[width=1.0 \columnwidth]{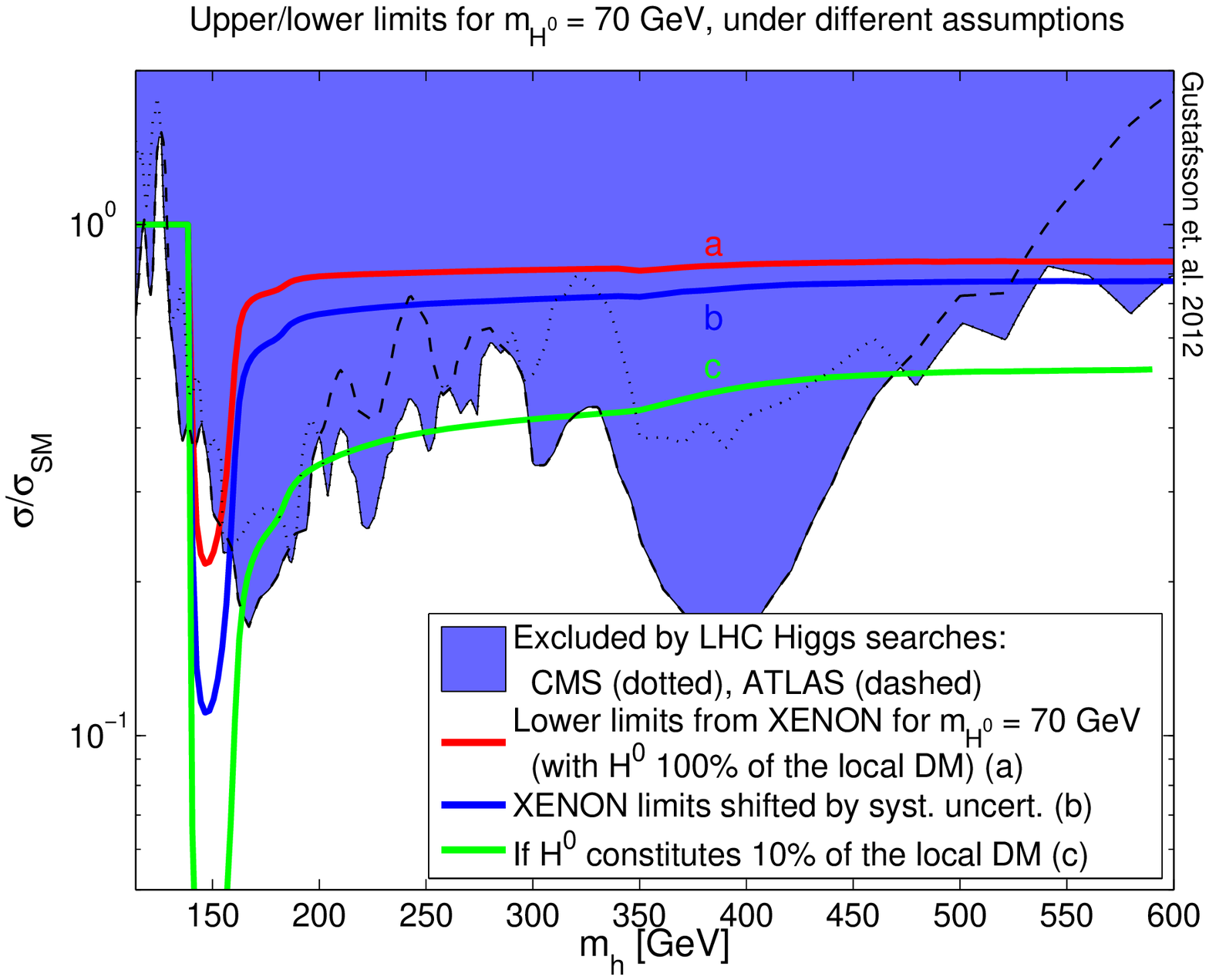}
\end{tabular}
\caption{
{\bf Left}: The solid lines show, for different $H^0$ masses, the lowest limits on the Higgs production rates $\sigma/\sigma_{\text SM}$ that the IDM can have at the LHC and still be compatible with XENON-100 data. These limits correspond to when none of the produced IDM states contribute to a signal in the LHC's Higgs boson searches, {\it i.e.}\  $\mathcal{R}=\mathcal{R}^\text{cons}$. The constraint from a thermal freeze-out calculation of $H^0$ has not been applied yet, and $H^0$ is assumed to make up the local DM density. All other constraints, EWPT and LEP limits in particular, are taken into account. The dark (blue) region is excluded by the current Higgs search results from the LHC.  Thus, only the white regions above the lines remain allowed in the IDM.  {\bf Right}: All lines assume an $H^0$ mass of 70 GeV. The upper (red) line assumes that $H^0$ particles provide all the local DM, the dark (blue) line below is after systematic uncertainties (see text) are included to diminish the direct detection constraints, while the lower (green) line applies when assuming that $H^0$ contribute only 10\% to the local DM density.}\label{fig:higgsbranch}
\end{figure*}

The latest results are based on analyses of $\sim 5$ fb$^{-1}$ of integrated luminosity. The CMS experiment set the strongest (preliminary) constraints on large Higgs masses until March 2012, excluding a SM Higgs boson over the mass range 127-600 GeV to 95\% CL, when all search channels are combined ($4.6-4.7$ fb$^{-1}$ of integrated luminosity) \cite{Chatrchyan:2012tx}. At that time, ATLAS presented their (preliminary) limits on large Higgs masses using up to 4.9 fb$^{-1}$ \cite{atlas-combined-update}. The CMS collaboration also updated their limits in some channels for $4.6-4.8$ fb$^{-1}$ \cite{cms-combined-update}. 
We will here use both the experiments' current best exclusion limits on a Higgs signal $\sigma/\sigma_{\text{SM}}$. Here $\sigma/\sigma_{\text{SM}}$ denotes the signal rate in units of the expected SM Higgs production cross section $\sigma_\text{SM}$. The 95\% CL upper limits, for all channels combined but for each experiment individually, will be used. In Figure~\ref{fig:higgsbranch} the excluded signal strength, as a function of the Higgs boson mass, is shown as the  blue (gray) region. The exclusion region represents the strongest of the two limits from the CMS (dotted line) and the  ATLAS (dashed line) experiments.

\subsubsection{Reduction of the Higgs signal in the IDM}
In the IDM, the new contributions to the SM-like Higgs width $\Gamma_h$ can have a significant impact on the LHC Higgs searches by effectively reducing the Higgs production cross section into SM particles. Since $H^0$ is neutral and stable, the Higgs decays into $H^0$ pairs will necessarily contribute to an invisible width. However, let us emphasize that the Higgs can also decay into $A^0$ and $H^\pm$ pairs which would further increase the Higgs width. The latter processes give rise to the production of (off- or on-shell) $Z$ and $W$ bosons that can make them partly visible in the Higgs search channels. 

Nevertheless, the exclusion limits on the Higgs mass range could very well be evaded within the IDM. The processes $h\rightarrow H^0H^0$, $h\rightarrow A^0A^0$ and $h\rightarrow H^+H^-$ enhance the Higgs decay width by:
\bea
\hspace{-2mm} 
\Delta\Gamma_{\rm IDM}&=&\frac{v^{2}}{16\pi m_{h}}\left[  
       \lambda_{H^0}^{2}\left(1-\frac{4m_{H^0    }^{2}}{m_{h}^{2}}\right)  ^{1/2}+ \right. \cr
&&\hspace{-15mm} + \left. 
       \lambda_{A^0}^{2}\left(1-\frac{4m_{A^0    }^{2}}{m_{h}^{2}}\right)  ^{1/2}+
     2\lambda_{3    }^{2}\left(1-\frac{4m_{H^\pm}^{2}}{m_{h}^{2}}\right)  ^{1/2}\right],   
     \label{eq:Gamma}
\eea
where $\lambda_{H^0,A^0,3}$ are given in Eq.~\refeq{eq:potential} and \refeq{eq:masses}.

In the narrow-width approximation, the signal strength $\sigma/\sigma_{\text{SM}}$, or equivalently the reduction factor $\mathcal{R}$, for producing SM particles $x\bar x$ is given by
\begin{eqnarray}
\;\;\;\mathcal{R}&=& \frac{\sigma_{\rm IDM}(pp\rightarrow h) \; \mathrm{Br}(h\rightarrow x\bar{x})_{\rm
    IDM}}{\sigma_{\rm SM}(pp\rightarrow h) \; \mathrm{Br}(h\rightarrow
  x\bar{x})_{\rm SM}}  \label{eq:RBr} 
 \label{eq:R} \\
&=& 
 \frac{\Gamma_{\rm SM}}{\Gamma_{\rm SM}+\Delta\Gamma_{\rm
     IDM}}\frac{\Gamma^{\rm SM}_{ h\rightarrow x\bar{x}}
       \, +\sum_{\phi}  \epsilon_{\sss \phi} \Gamma_{ h\rightarrow \phi^\dag \phi}}
     {\Gamma^{\rm SM}_{h\rightarrow x\bar{x}}}, \nn
\end{eqnarray}
where the sum runs over $\phi=A^0, H^\pm$ and $\Gamma_\text{SM}$ is
the total decay width of the Higgs in the SM. In the last step we use the fact that the Higgs production and decay rates into SM particles $x\bar x$ are unchanged to first order\footnote{For an effect at the loop level, see {\it e.g.}\ the study in \cite{Arhrib:2012ia}  of the $\gamma\gamma$ channel.} ($\Gamma^{\sss \rm SM}_{\sss h\rightarrow \rm SM} = \Gamma^{\sss \rm IDM}_{\sss h\rightarrow \rm SM}$). $\epsilon_\phi$ denotes the efficiency with which $A^0A^0$ and $H^+H^-$ may contribute to the current $x\bar x$ Higgs search. This efficiency may be expected to be low due to the fact that the final states will contain extra invisible $H^0$ states and therefore, in principle, have different characteristics than the pure SM $x\bar x$ final states. 

This means that for the whole range of Higgs masses, even if excluded within the SM, the LHC limits could potentially be evaded within the IDM.

In the next subsection we will argue that, for the models of interest for our study, Higgs decay into all IDM particles will effectively be invisible. In that case the reduction factor in Eq.~\refeq{eq:R} reduces to
\beq
\mathcal{R}^\text{cons} = \frac{\Gamma_{\rm SM}}{\Gamma_{\rm SM}+\Delta\Gamma_{\rm IDM}}.
\eeq
In general, the IDM contribution could, in principle, also enhance certain SM Higgs signatures, depending on the specific model and search channel. However, the effect of such a contribution would only give stronger exclusion limits on large Higgs masses than in the SM. Taking $\mathcal{R} = \mathcal{R}^\text{cons}$ is thus the most conservative choice when it comes to determining to what extent the IDM is excluded and therefore the one that we will adopt in the following.

\subsubsection{Higgs searches and the IDM}
The $WW$ and $ZZ$ search channels are the most effective ones in the search for heavy Higgs bosons, and below we list their most sensitive subchannels. We quote the excluded SM Higgs masses, as this indicates where the searches could be sensitive enough to exclude Higgs masses in the IDM.

\begin{itemize}
\item 
$h\rightarrow ZZ^{(*)}\rightarrow 4l$ with $l=$ \{electron, muon\}. By these lepton channels alone, the CMS experiment excluded at 95\% CL SM Higgs boson masses in the ranges 134-158, 180-305 and 340-465 GeV \cite{Chatrchyan:2012dg}. At the same confidence level, ATLAS excluded the ranges 134-156, 182-233, 256-265 and 268-415 GeV \cite{ATLAS:2012ac}. An important requirement in all these searches is that at least one same-flavor opposite-sign lepton pair has an invariant mass in a window around the $Z$ mass. 
\item 
$h\rightarrow ZZ\rightarrow 2l2\nu$. In this channel, the events are required to contain a minimum amount of missing transverse energy and the lepton pair is required to form an on-shell $Z$ boson. The CMS collaboration was able to use this channel alone to exclude SM Higgs masses in the range 270-440 GeV at 95 \% CL \cite{Chatrchyan:2012ft}, and the corresponding range excluded by ATLAS is 320-560 GeV \cite{:2012va}.
\item
$h\rightarrow ZZ\rightarrow 2l2q$. This search requires the invariant mass of the jet pair to correspond to an on-shell $Z$ boson. ATLAS excludes SM Higgs masses in the ranges 300-310 and 360-400 GeV at the 95\% CL~\cite{:2012vw}, while CMS could not exclude SM Higgs production cross sections by the use of this channel alone \cite{Chatrchyan:2012sn}.
\item 
$h\rightarrow WW\rightarrow l\nu l\nu$. In this channel, the events are required to contain at least two leptons of opposite sign and missing transverse energy. Cuts on the transverse mass, reconstructed from the lepton pair together with the missing transverse energy, are also applied. The CMS excludes at 95\% CL a SM Higgs mass in the range 129-270 GeV \cite{Chatrchyan:2012ty}, and ATLAS the range 130-260 GeV \cite{Collaboration:2012sc}. 
\end{itemize}

Higgs decays into $H^0$ would be invisible, but might it also be the case that decays into $A^0$ and $H^\pm$  escape detection in the above search channels?

The decay channel $h\rightarrow A^0A^0$ would give rise to two $Z$ bosons and could be visible in the above $ZZ$ search channels. It would however only give a visible contribution if $(m_{A^0}-m_{H^0})$ is large enough to produce on-shell $Z$ bosons via the decay $A^0\rightarrow H^0+Z$.\footnote{Even in the case of on-shell $Z$'s, the characteristics of the final states are altered by the presence of $H^0$'s giving rise to $\ET$. In the $ZZ\rightarrow 4l$ channel this would lead to a smearing of the $4l$ invariant mass spectrum, thereby evading a peak search, but could potentially contribute to the observation of a less constraining broad excess. 
}

In the $WW\rightarrow 2l2\nu$ Higgs search channel, the final state is required to include two opposite-sign leptons and missing energy.
The $h\rightarrow A^0A^0$ and $h\rightarrow H^+H^-$ production, with the subsequent decays $A^0\rightarrow H^0 + Z$ and $H^\pm\rightarrow H^0+W^\pm$, could pass these requirements and one can imagine that this could contribute to a signal in this search channel. Let us therefore take a closer look at this possiblity, to see if the contribution could be significant. So far this channel only excludes SM Higgs masses in the range 130-270 GeV, and we therefore expect that it is only within this same mass range that Higgs bosons can be excluded in the IDM. This statement is motivated by the use of cuts on the transverse mass, that sets the SM Higgs mass for which the limit applies. This `transverse mass' variable corresponds to the Higgs boson mass in the SM and should roughly do so also in the IDM. This entitles the use of the same $\sigma/\sigma_{SM}$
limit for the Higgs in  the IDM as in the SM.

In this specific mass range, Higgs decays into $A^0A^0$ and $H^\pm H^\mp$ will however never contribute to the $WW\rightarrow 2l2\nu$ Higgs search channel. This is because for $m_h\lesssim$ 160 GeV the LEP limits \cite{Pierce:2007ut,Lundstrom:2008ai} already exclude almost all inert particles $A^0$ and $H^\pm$ with masses less than 80 GeV, which are the only masses that could have been kinematically accessible for these Higgs decays. The exception, with lighter $m_{A^0,H^0}$, occurs only when the mass splitting $m_{A^0}\! -\! m_{H^0}$ is very small, and the final-state fermions are then too soft to contribute. Moreover, in the region $160\;{\rm GeV}\! <\!m_h\!<\!270\;{\rm GeV}$ it turns out that IDMs which account for all the DM are excluded irrespective of whether the $A^0$ and $H^\pm$ states are invisible to the Higgs searches or not (see Figure~\ref{fig:higgsbranch}).

\smallskip
This means that for many models, in particular those that have a mass difference $(m_{A^0}-m_{H^0})$ too small to produce $Z$ bosons on shell, the IDM contributions to the Higgs width can be treated as invisible in the current LHC searches for heavy Higgs bosons. Our arguments for such a treatment were based on the channels important for the searches in the high $m_h$ region, while for low Higgs masses, other channels could be more important. Nevertheless, we will apply the same assumption to all our models as this will not alter our discussion.

\subsection{Constraints on IDM from LHC and XENON-100 combined}
Figure \ref{fig:higgsbranch} shows the LHC Higgs exclusion limit together with IDMs that have the largest invisible Higgs width possible and still pass \mbox{XENON-100} direct detection constraints. As we can conclude from the above discussion, all the inert states resulting from Higgs decay can be regarded as effectively invisible when the mass difference $(m_{A^0}-m_{H^0})$ is less than $m_Z$, {\it i.e.}\ $\mathcal{R}=\mathcal{R}^\text{cons}$. In Figure~\ref{fig:higgsbranch}, we present lines for when we take $\mathcal{R}=\mathcal{R}^\text{cons}$ for some representative $m_{H^0}$ masses.

Once $m_{H^0}$ and $m_h$ are fixed, Eq.~\refeq{eq:SSI} and the XENON-100 exclusion limit on $\sigma^\text{SI}$ determine the largest available value of $\lambda_{H^0}$, and consequently $\Delta\Gamma_{h\rightarrow H^0H^0}$.  The largest values of $\lambda_{A^0}$ and $\lambda_3$, driving the two other contributions to $\Delta\Gamma_\text{IDM}$ in Eq.~\ref{eq:Gamma}, can be found numerically under the imposition of all the other IDM constraints listed in Sec.~\ref{sec:2}. The only exception is that we do not yet impose that $H^0$ accounts for the total WMAP DM relic abundance. Instead, we immediately assume that the local $H^0$ density provides the observed DM density that is relevant for the constraints on DM direct and indirect detection. This is in order to keep the discussion more general at this stage, and not include constraints from the freeze-out process occurring in the early Universe. We notice that the LEP and EWPT bounds give the most crucial limits to constrain $\lambda_{A^0}$ and $\lambda_3$ after the XENON bound has been imposed. Together with the XENON and LHC constraints, they are efficient in excluding IDMs with heavy Higgs masses.

We see that even without including the relic density calculations, the XENON and LHC Higgs searches, if taken at face value, exclude most of the IDM scenario with large Higgs masses. Only two exceptions appear -- see the left plot in Figure~\ref{fig:higgsbranch}. 

\begin{figure}[t!]
\includegraphics[width=1.00 \columnwidth]{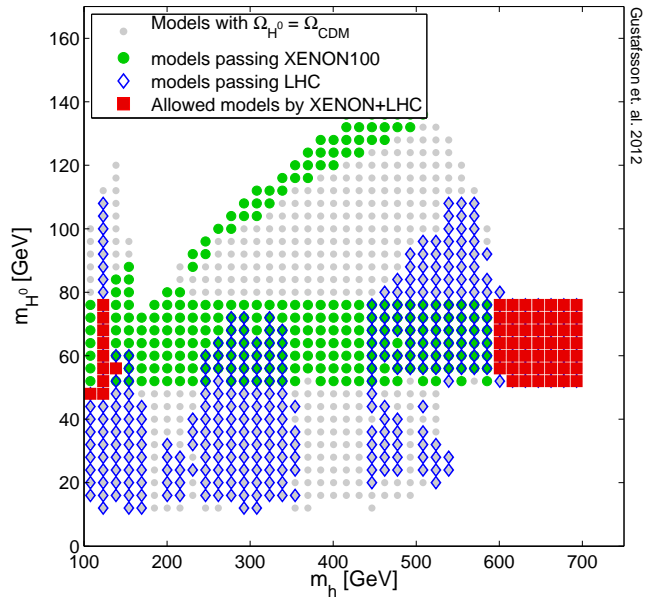}
\caption{All the points correspond to IDMs giving rise to a DM relic density in agreement with WMAP. Small (grey) points represent models that neither pass the LHC Higgs searches nor XENON-100 constraints, diamonds (in blue) represent models that pass LHC and large (green) points represent models that pass XENON-100 constraint individually. Squares (in red) represent models that pass the constraints from both the LHC and XENON-100 experiments. All the other constraints of Sec.~\ref{sec:2} have also been taken into account. } \label{fig:AvailableRegion}
\end{figure}

First, we have the low mass WIMP, with {\it e.g.}\ $m_{H^0}=8$ GeV, which could give rise to large Higgs decay branching ratio into $H^0$. As discussed in Sec.~\ref{sec:dd}, this case is already excluded by \mbox{XENON-10} and Fermi-LAT data and is presented for illustration only. The second exception arises in the large mass region for $m_{H^0}\sim80-150$~GeV, which might still be viable for the largest Higgs masses. 
However, if we take into account also the constraint from having the DM candidate, $H^0$, as a thermal relic, this region is no longer allowed. This is clearly seen in Figure~\ref{fig:AvailableRegion} where the relic density calculation has been included. We are thus able to exclude the so-called 'new viable region' of IDM found in \cite{LopezHonorez:2010tb} even before direct detection experiments have fully probed this regime of the IDM. Therefore none of these exceptions provides good models. 

Also the possibility to have models with Higgs masses above 600 GeV still remains. The LHC has only presented bounds on  $\sigma/\sigma_\text{SM}$ for Higgs masses below 600 GeV, and we can therefore not use this quantity directly to exclude models with such large Higgs masses. The EWPT and unitarity constraints, however, limit the Higgs mass to be below $\sim700$ GeV (also the triviality/perturbativity bound would disfavor larger Higgs masses \cite{Lindner:1985uk, Wingerter:2011dk}). As can be seen from Figure~\ref{fig:AvailableRegion}, when the thermal relic density calculation has been included, the DM mass range $m_{H^0} \sim 45 - 80$ GeV with a very heavy Higgs in the range 600-700 GeV is still an allowed region.

In Figure~\ref{fig:AvailableRegion} we present the result of a random scan in the $m_{DM}\, \in \, [15-170]$ GeV parameter space of the IDM giving rise to an $H^0$ relic abundance in agreement with WMAP \cite{Larson:2010gs} at the 3$\sigma$ level. All the constraints from Sec.~\ref{sec:2} are now included. The plot illustrates in the $m_h-m_{H^0}$ plane the IDMs that pass the constraints set by XENON-100, the LHC Higgs searches and WMAP. We see that many models pass either the direct detection or the LHC Higgs bounds individually. In the heavy Higgs region there are no surviving models,  except for the region $m_h \gtrsim 600$ GeV and $m_{H^0} \sim 45-80$ GeV (see also the plots in Figure~\ref{fig:higgsbranch}). We thus conclude that in order to have an IDM that makes up all the DM and has a SM-like Higgs boson in the 160-600 GeV mass range, at least one of our imposed constraints has to be relaxed.

\subsection{Accommodating a heavy Higgs boson \\and DM in the IDM} \label{sec:4B}
One of the original motivations for studying the IDM was that it could alleviate the LEP paradox in the SM by allowing for a heavier Higgs particle while staying in agreement with EWPT. We have shown above that constraints from direct detection in combination with the SM Higgs search essentially rule out large Higgs masses up to  $\sim 600$ GeV in the IDM. 

In this section, we investigate the assumptions that could be relaxed in order to allow for a large range of high Higgs masses ($m_h>160$ GeV) within the IDM.  In particular, we will allow for larger values of $\lambda_{H^0}$ by suppressing the bound that derives from direct detection searches. In that way, models with larger invisible Higgs branching ratios will become available, which consequently give lower signal strengths in the LHC Higgs searches. This is illustrated in the right panel of Figure~\ref{fig:higgsbranch}.

The bound on $\lambda_{H^0}$ can be suppressed in two ways:
\begin{enumerate}
\item by assuming that the DM from IDM does not account for the entire DM abundance: the green line in the right panel of Figure~\ref{fig:higgsbranch} 
assumes that $H^0$ constitutes only $10\%$ of the local DM density $\rho_0$. This suppresses the constraint on $\sigma_{H^0-p}^\text{SI}$ by the same factor.
\item by considering systematic uncertainties in direct detection: the dark blue lines in the right panel of Figure~\ref{fig:higgsbranch} take into account a smaller form factor $f=0.26$ \cite{Ellis:2008hf,Mambrini:2011ik}, a smaller local DM density $\rho_0=$ 0.2 GeV/cm$^{-3}$ \cite{Green:2011bv}, and, in addition, include a minor effect of 10\,\% weakening of the XENON-100 cross section limits due to uncertainties in the local WIMP velocity distribution \cite{Green:2011bv}. Concerning the local DM density, there have been recent improved measurements constraining it to the range $\rho_0= 0.3\pm0.1$ GeV/cm$^{-3}$ \cite{Bovy:2012tw} (see also \cite{Bidin:2012vt}).
\end{enumerate}

These reconsiderations weaken the constraints on  $\lambda_{H^0}$, and IDMs with $m_{h} \gtrsim 500$ GeV could be allowed. Also $m_{h}$ around 320 GeV could be allowed if only the LHC constraints from CMS are considered. However, the preliminary analysis recently presented by ATLAS  \cite{atlas-combined-update} does not show any excess  around $m_h=320$~GeV as CMS does, but instead puts very strong constraints in the  300 to 450~GeV mass range. There are also uncertainties related to the absolute calibration of  cross section limits at the LHC on $\sigma/\sigma_{\text{SM}}$. We choose here not to take into account such potential additional uncertainties.

\medskip
If $H^0$ particles constitute only a fraction of the DM density they would more easily pass direct detection constraints (now rescaled by $\Omega_{\rm CDM}/\Omega_{H^0}$) while having a larger $\lambda_{H^0}$ coupling, and then be able to evade the LHC Higgs limits. It then remains to be shown if such models exist that have such a low relic density while not exceeding the other constraints in Sec.~\ref{sec:2}. The possible mechanisms for this in the IDM are as follows \footnote{Models with annihilation dominantly into fermions have $\langle \sigma v\rangle \propto \lambda_{H^0}^2$, and are already in the region excluded by direct detection searches. This can be seen in Figure~\ref{fig:dd}, where $m_{H^0}\lesssim 40$ GeV corresponds to models having annihilations into fermions only. In that framework, increasing $\lambda_{H^0}$ would not alter the bounds from direct detection searches. Indeed, these bounds derive from the quantity \mbox{$\sigma^{SI} \!\times\! \Omega_{H^0}  \propto \lambda_{H^0}^2/\langle \sigma v\rangle$} which is unchanged under a rescaling of $\lambda_{H^0}$.}:
\begin{itemize}
\item {\it Annihilation via $h$ at the resonance} ($m_{H^0}\sim m_h/2$):\\ In the case of a heavy Higgs boson, the resonance could only occur when $m_{H^0}\gtrsim 80$ GeV and annihilations into gauge bosons already provide an efficient annihilation mechanism.
\item {\it Coannihilations} ($m_{H^0}\sim m_{A^0}$ or $m_{H^0}\sim m_{H^\pm}$ ):\\ This is relevant for small mass differences when $m_{A^0,H^\pm}/m_{H^0}\lesssim 1.1$. For large Higgs masses, the EWPT also requires that $(m_{H^\pm}-m_{A^0})\times(m_{A^0}-m_{H^0})$ is positive \cite{Barbieri:2006dq}. This means that $m_{H^\pm}>m_{A^0}$ and that the mass difference between the two neutral inert scalars has to be small. For the tetralepton search channel that we will investigate in the next section, this has the implication that the leptons from the decay $A^0\rightarrow H^0$ are too soft to be detected at the LHC.
\item {\it Annihilation to} $WW, ZZ \;\mathrm{{\it and}}\; t\bar t$ ($m_{H^0}\gtrsim$ m$_W$):\\ Strong annihilation channels into gauge bosons become kinematically available already for $m_{H^0}$ just below $m_W$, $m_Z$ or $m_t$.
\end{itemize}

Although all three of the above mechanisms could be viable, we will in the next section only consider models where the relic density is suppressed by the last type of mechanism. This is because we want to investigate the best prospects for detecting the IDM in the tetralepton channel at the LHC, and the simplest scenario to consider is then when the $WW$ annihilation channel regulates the DM abundance.

We will also consider benchmark models that give a relic density in agreement with 100 \% of the observed DM. However, for these models systematic uncertainties for the direct detection searches have to be included, as described above, to make them pass all constraints.

\section{The multilepton signal}   \label{sec:4}
The inert scalars can only be produced in pairs, since each inert particle has negative $\mathbb Z_2$ parity contrary to the SM particles. At tree-level, the relevant hard processes producing final states with four leptons or more, are via the gauge bosons and the Higgs boson:
\bea
 q\bar q'   &\rightarrow&  W^{\pm} \rightarrow A^{0}H^{\pm} \label{eq:qq1}\\
 q\bar q'   &\rightarrow&  Z/\gamma/h \rightarrow H^{+} H^{-}.   \label{eq:qq2}
\eea
The tree-level contribution to $ q\bar q'  \rightarrow  h \rightarrow A^0A^0$ is negligible but at loop level, gluon fusion into Higgs is important for $A^0A^0$ and $H^{+} H^{-}$ production.

\begin{figure}[t] 
\includegraphics[width=0.47 \columnwidth]{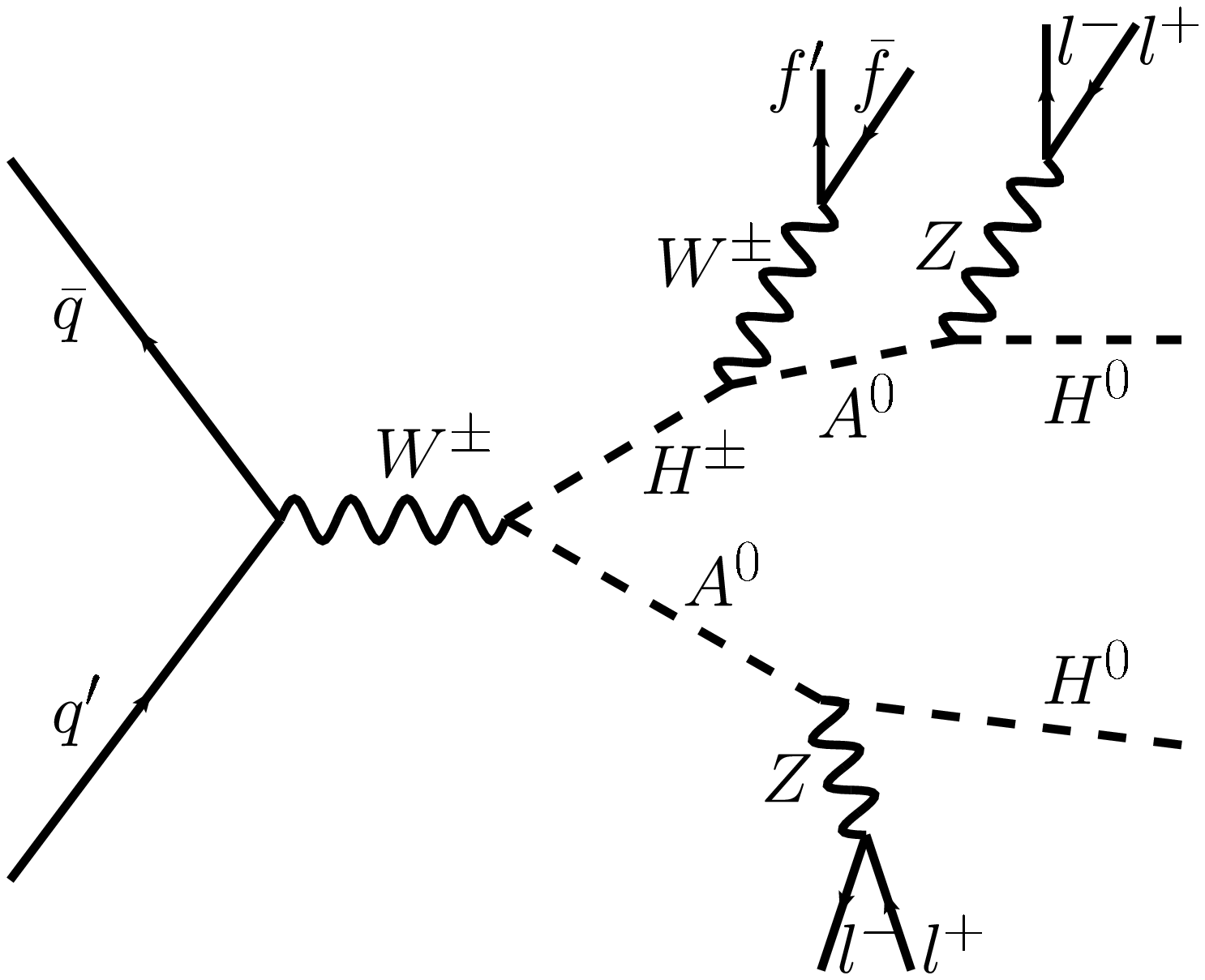}
\hspace{\stretch{1}} \vspace{\stretch{1}}
\includegraphics[width=0.47 \columnwidth]{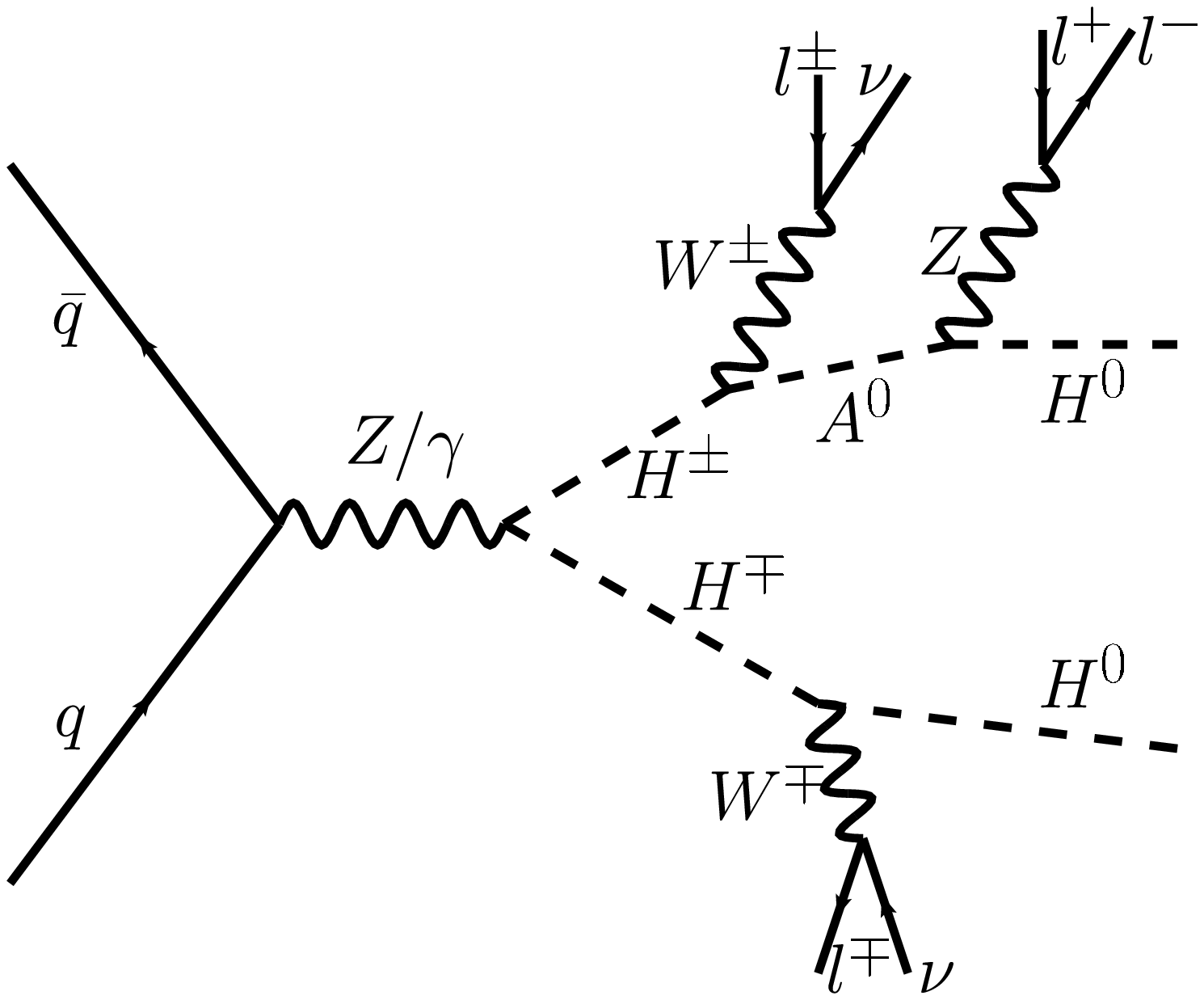}

\includegraphics[width=0.47 \columnwidth]{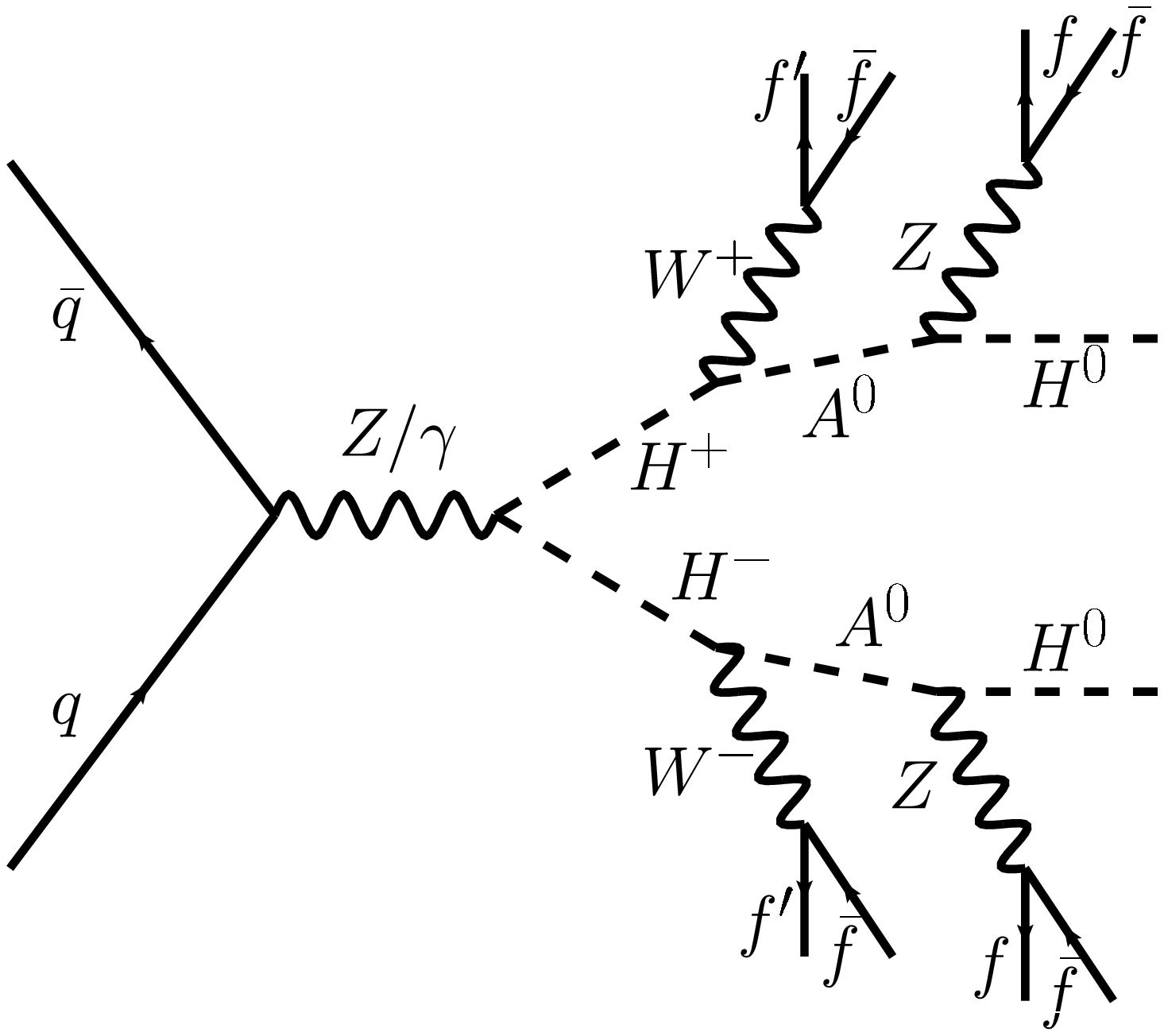}
\hspace{\stretch{1}} \vspace{\stretch{1}}
\raisebox{15pt}{\includegraphics[width=0.47\columnwidth]{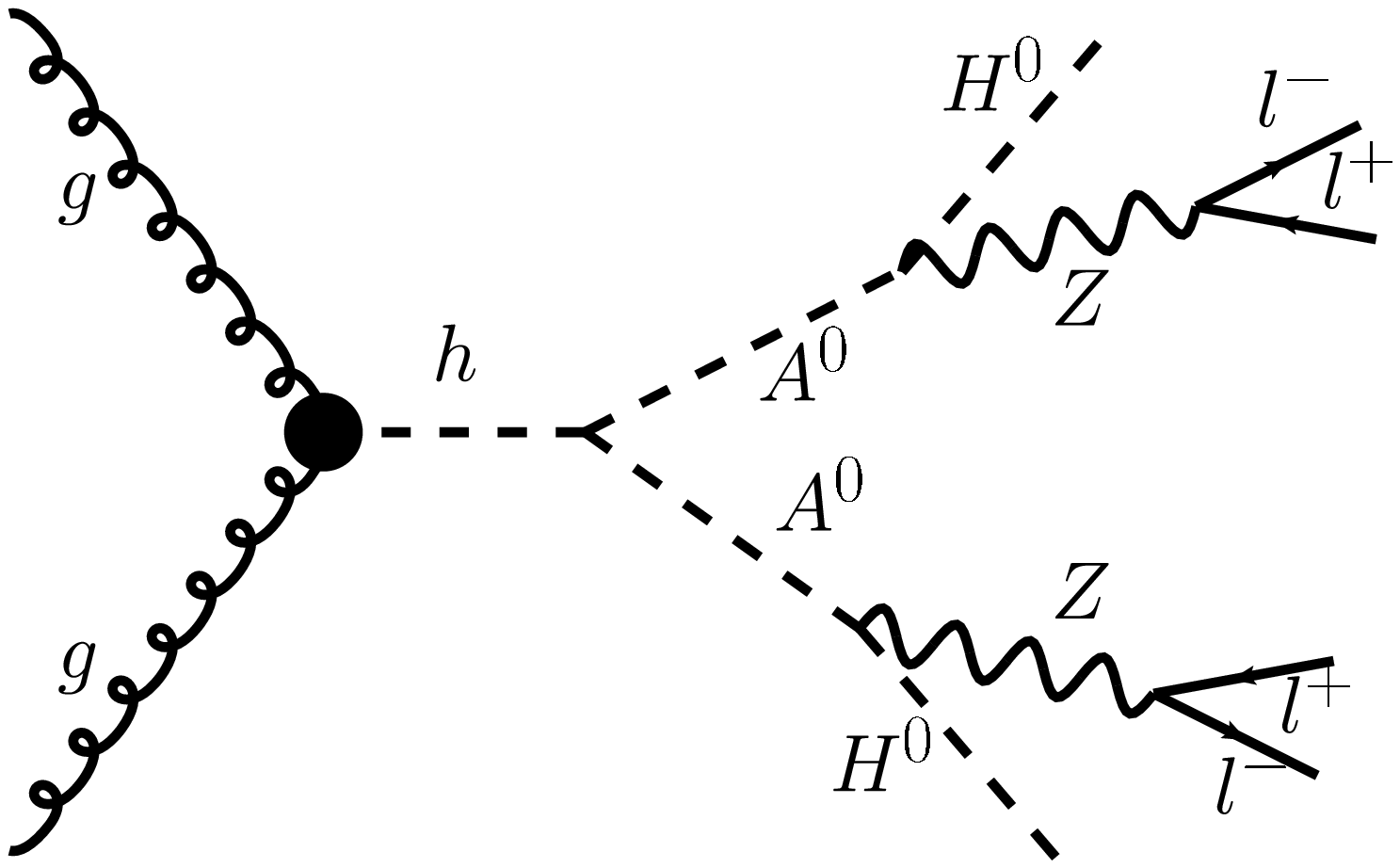}}

 \caption{Feynman diagrams contributing to $pp\rightarrow 
 4l+ \displaystyle{\not}E_T$ in the IDM. } 
 \label{fig:1} 
\end{figure}

After the inert particles are produced, they will cascade decay through the processes:
\bea 
\label{eq:HpA}
   &H^{\pm} \rightarrow  
   \left\{\begin{array}{ll} H^{0} W^{\pm}  \\A^{0} W^{\pm}\end{array}\right. 
   &\mathrm{,\; and}\;\;  A^{0}   \rightarrow H^0 Z   
\\ \mathrm{or}&&\nn \\
\label{eq:HpA2}
  &H^{\pm} \rightarrow H^0 W^{\pm} 
  &\mathrm{,\; and }\;\;  
  A^{0}   \rightarrow \left\{ \begin{array}{ll} H^{\pm} W^{\mp}\\H^0Z\end{array} \right., 
\eea
depending on whether $H^\pm$ or $A^0$ is the most massive inert state. The gauge bosons will, with their respective branching ratios, decay into fermions, $f$, according to
\bea 
W^{\pm} \rightarrow f^\pm \nu &\mathrm{and}&   Z \rightarrow f^\pm f^\pm.
\eea
Figure~\ref{fig:1} illustrates these production and decay chains. Our focus will be on the production of four or more leptons, $l$ (which in this context refers only to electrons and muons), where the SM background is expected to be very low.

The cross sections and decays widths will be calculated using \mg/\me \cite{Alwall:2007st}, and their streamlined interface with \pythia \cite{Sjostrand:2006za} and \pgs \cite{PGS} to simulate hadronization and detector response. To be able to generate signal events in practice, we split the processes into separable steps, in order to diminish the phase space from the otherwise up to ten-body final-state processes. In the first step the inert states are produced on shell, as in Eqs.~\refeq{eq:qq1}-\refeq{eq:qq2}. In the following steps, {\it i.e.}\ Eqs.~\refeq{eq:HpA}-\refeq{eq:HpA2}, the inert scalar particles are also taken to be on shell while keeping the virtuality of the gauge bosons fully general. As a check of the validity of this approximation, we note that the inert particles' resonances for our benchmark models are indeed narrower than gauge bosons'. In all cases, the width of the $A^0$ is small, of the order $10^{-4}-10^{-5}$ GeV, due to the small mass difference to $H^0.$\footnote{This does not, however, make $A^0$ sufficiently long lived to give rise to displaced vertices.} The width of $H^\pm$ varies more, but is still smaller than the $W$ width for all our models except one, which anyway has mass differences that allow both $W$ and the inert state to be on shell simultaneously. 
Moreover, the most important contribution will come from direct production of $A^0$ pairs, and the models for which the production of $H^\pm$ gives a significant contribution to the signal coincides with large enough $\Delta m_{H^\pm A^0}=m_{H^\pm}-m_{A^0}$ to allow $A^0$ to be on shell in Eq.~\refeq{eq:HpA}.

\subsection{Production of inert scalars via gauge fields}
\begin{figure}[t] 
\begin{overpic}[height=8cm, trim = 0 0 0 0, clip=true]{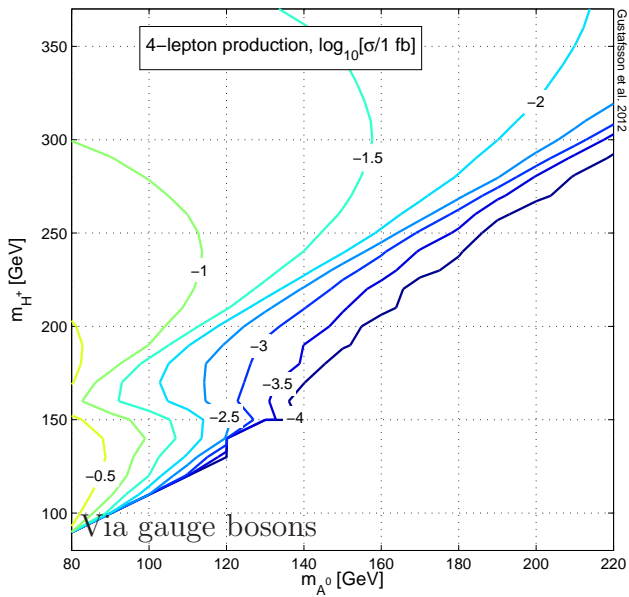}
\put(30,25){\large Via gauge bosons} 
\end{overpic}
  \caption{Cross sections $\sigma$ into four or more leptons from $H^+H^-$ and $H^\pm A^0$ production \emph{via gauge bosons} (in units of 10 logarithms of $\sigma$ in fb). Here $m_{H^0}=70$ GeV. This contribution to the cross section is independent of the Higgs mass.}
 \label{fig:4} 
\end{figure} 

In this subsection, we discuss the general expectations of the $\geq\!4$ lepton signal strength from inert scalars produced via gauge bosons. Some of the contributing diagrams are shown in the first three panels of Figure~\ref{fig:1}. As the gauge couplings are fixed, the production cross sections of the heavier inert states are fully determined by their masses, and their decay patterns by their mass splittings:
\bea
\Delta m_{H^\pm H^0}  &=&  m_{H^\pm}-m_{H^0},\\
\Delta m_{H^\pm A^0}  &=& m_{H^\pm}-m_{A^0},\\ 
\Delta m_{A^0 H^0}      &=&  m_{A^0}-m_{H^0}.
\eea

The processes in Eq.~\refeq{eq:qq1}-\refeq{eq:qq2} can give rise to final states with four or more leptons if the mass hierarchy is $m_{H^0}<m_{A^0}<m_{H^\pm}$, and this is the mass hierarchy  we will consider in this section. For large Higgs masses ($m_h\gtrsim$ 160 GeV),  the EWPT constraints require $H^\pm$ to be the heaviest state and the mentioned mass hierarchy is  then just a consequence of $H^0$ being the DM candidate. Apart from that, the Higgs mass has no impact on our results in this section. 

As we will see, even for optimal parameter values, the gauge-mediated contribution to a four-lepton signal in the IDM will not be enough to render the model detectable. Here we merely study under what conditions the contribution from gauge mediated production can become non-negligible, and we will turn to the more significant contribution from gluon fusion in the next section.

As our interest is in the detection of leptons, the branching ratios $A^0\!\rightarrow\! H^0 \ell^+\ell^-$ and $H^\pm\!\rightarrow\! A^0 l^\pm\nu$ are important.  For very small mass splittings $\Delta m_{A^0 H^0}$ the Br($A^0\!\rightarrow\! H^0 \ell^+\ell^-$) can be large, but give rise to leptons that are too soft to be isolated. For increased mass splitting, decay modes into the more massive quarks open up, and the branching ratio into leptons decreases, approaching 6.7\,\% which is the result for an on-shell $Z$ boson. 
 A small $\Delta m_{A^0 H^0}$ also gives larger Br$(H^\pm\!\rightarrow\! A^0 W^\pm)$ as a large mass splitting would kinematically favor decay into $H^0$; especially if $W$ becomes on shell. Again a small mass shift becomes weighed against the ability to produce hard enough leptons for detection. 
For a fixed $\Delta m_{A^0 H^0} $, increasing $m_{H\pm}$ will typically increase Br$(H^\pm\!\rightarrow\! A^0)$, but at the cost of lowering the production cross section of heavier $H^\pm$.

In Figure~\ref{fig:4} we show the cross section for the gauge mediated contribution to the production of four or more leptons. We calculate the tree-level cross sections with \mg/\me and apply a corrective factor, a so-called K-factor, of 1.2 to achieve agreement with the NLO results in \cite{Cao:2003tr,Aad:2009wy}.

\begin{table*}
    \begin{tabular}{@{\extracolsep{\fill}} c|c|c|c|c|c|c||c|c|c||c|c|c|c} 
   Benchmark 	&$m_{h}$	&$m_{H^0}$	& $m_{A^0}$	&$m_{H^\pm}$	& $\mu_2^2$& $\;\;\;\lambda_2^\mathrm{min}$\;\;\; & $\;\;\;\lambda_{H^0}$\;\;\; & $\lambda_{A^0}$ & $\lambda_{H^\pm}$ &$\sigma v_\mathrm{tot / 3-body}$&$\sigma v_\mathrm{\gamma\gamma / \gamma Z}$&$\sigma^{\text{SI}}\times\frac{\Omega_{H^0}}{\Omega_{\text{CDM}}}$&$\Omega_{H^0} h^2$	\\ \hline
    IDM-A1    	& 300	& 72.0	& 110	& 210       & $ 72^2$			&0				&	0			& 0.22	& 1.3	 	&1.5 / 1.5			&7.9 / 8.0 			& $4\cdot10^{-5}$			&0.108 	       	\\ 
    IDM-A2    	& 500    	& 71.8	& 110  	& 230 	& $ 72^2 $		&$10^{-7}$		&$-10^{-3}$		& 0.22	& 1.6	 	&1.5 / 1.5			&7.9 / 8.0 			& $5\cdot10^{-5}$			&0.111 \\ \hline
    IDM-B1    	& 320  	& 77.5	& 105	& 152	& $-1.6\cdot10^{4}$	&0.080			&	0.70			& 0.86	& 1.3	 	& 1.1/ 0.83		& 2.8 / 2.5			& 16						& 0.110	\\ 
    IDM-B2    	& 550  	& 76.0	& 140	& 220	& $-6.0\cdot10^{4}$	&0.38			&	2.1			& 2.5		& 3.4	 	& 1.2 / 0.95 		& 3.7 / 3.7			& 18						& 0.113	\\ \hline
      IDM-C1		& 320	& 91.0		& 120	& 190	& $-6.5\cdot10^4$	& 1.3			& 2.3				& 2.5		& 3.2		& 210 / 0			& 15 / 62 			& 0.13				& 0.00105			\\
      IDM-C2		& 280	& 81.0		& 130	& 190	& $-3.5\cdot10^4$	& 0.50		&	1.3			& 1.7		& 2.3		& 16 / 14			& 2.8 / 13			& 8.1					& 0.00979		\\
    IDM-C3    	& 550  	& 92.0	& 140	& 230	& $-1.6\cdot10^{5}$	&2.7				&	5.4			&  5.7	&  7.2 	& 19 / 0			& 0.38 / 3.6 		& 7.5					&0.0106	\\
    IDM-C4    	& 550  	& 85.9	& 140	& 230	& $-6.0\cdot10^{4}$	&0.38			&	2.1			& 2.5	 	&  3.8 	& 23 / 0			& 6.1 / 10			& 1.4					&0.0110	\\ 
      \end{tabular}
    \caption{Benchmark models. Masses in units of GeV and $\lambda_2$ (not directly relevant) is taken to its, by vacuum stability, minimal allowed value $\lambda_2^\mathrm{min}$.  Annihilation cross sections, at relative impact velocity $v\rightarrow 10^{-3} c$, are in units of $10^{-26}$ cm$^3$/s for $\sigma v_\mathrm{tot,3-body}$ and in units of $10^{-29}$ cm$^3$/s for $\sigma v_\mathrm{\gamma\gamma,\gamma Z}$. Spin-independent cross sections $\sigma^{\text{SI}}$ in units of $10^{-45}$ cm$^2$.}
    \label{tab:1}
\end{table*}

\subsection{Production of inert scalars via SM Higgs}\label{sec:gluon}
\begin{figure}[t] 
\begin{overpic}[height=8cm, trim = 0 0 0 0, clip=true]{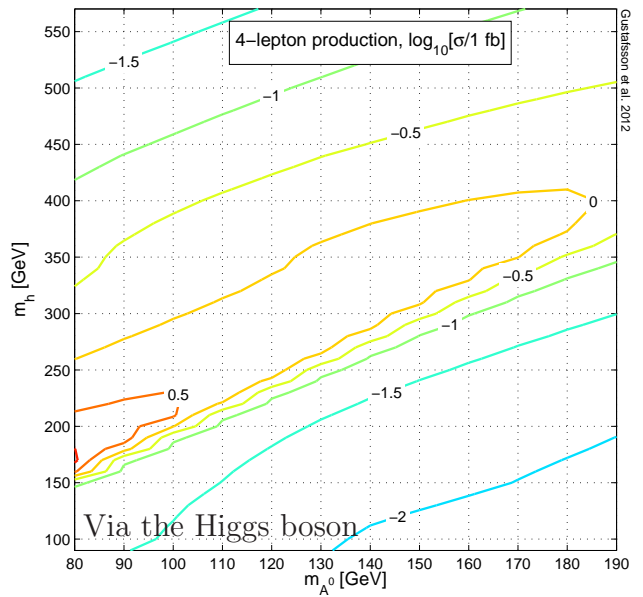}
\put(30,25){\large Via the Higgs boson}
\end{overpic}
  \caption{Cross sections $\sigma$ into four or more leptons from $A^0A^0$ and $H^+H^-$ production \emph{via the Higgs scalar} $h$ (in units of 10-logarithms of $\sigma$ in fb). In this plot $m_{H^0}=70$ GeV, $\mu_2^2=0$ and $m_{H^\pm}=220$ GeV. Only Higgs production via gluon
    fusion is included.}
 \label{fig:5} 
\end{figure} 

The SM Higgs production at LHC is dominated by gluon fusion -- dominantly induced by the loop of a top-quark coupled to the Higgs boson \cite{Djouadi:2005gi}. The couplings of inert particles to the Higgs can then give a significant contribution to the production of four leptons through the processes
\beq\label{eq:gg}
 gg  \rightarrow h \rightarrow A^0A^0, H^{+} H^{-}.
\eeq
In the $A^0A^0$ channel one obtains four leptons in the final states independently of the values of $m_{H^\pm}$ and Br$(H^\pm\rightarrow A^0)$. This process is shown in the last diagram of Figure~\ref{fig:1}. The signal strength will, apart from $m_{H^0}$ and $m_{A^0}$, also depend on $m_h$ and $\lambda_{H^0}$.  Unlike the processes considered in the previous section, the study of this process is strongly related to the SM Higgs search and to the search for DM in direct detection experiments. Given $m_{H^0}$ and $m_h$, direct detection data constrains the coupling $\lambda_{H^0}$ between $H^0$ and the Higgs boson, which for a given mass $m_{A^0}$ also limits the size of the Higgs coupling to $A^0$
\beq 
\lambda_{A^0}=\lambda_{H^0} +\frac{m_{A^0}^2-m_{H^0}^2}{v^2}.
\eeq

To generate $gg\rightarrow A^0A^0 ,H^{+} H^{-}$ events we make use of \mg/\me's implementation of the Higgs effective theory, where the Higgs boson couples directly to gluons. The effective coupling between the Higgs boson and the gluons depends on the Higgs mass and we match the cross sections obtained with \mg/\me to the next-to-next-to-leading order results for Higgs production via gluon fusion in the SM \cite{Dittmaier:2011ti}\footnote{In fact, the default effective operator implementation in \mg/\me-4.4.32 is not well suited for large Higgs masses. The deviation is as much as a factor 4.0, 4.6, 8.0 and 7.2 for $m_h =300,320,500,550$, respectively, compared to the results in \cite{Dittmaier:2011ti}.}.

At the largest Higgs masses the  vector boson fusion could also start to become relevant, but we are conservative in the sense that we do not include such, or other subdominant, Higgs production contributions to our IDM signal. In Figure~\ref{fig:5} we show the IDM cross section to four or more leptons by the Higgs mediated interactions in Eq.~\refeq{eq:gg}.

\subsection{Background} 
The requirement of leptons in the final state enables a signal to be extracted from the otherwise huge QCD background at hadron colliders. In order to simulate the SM background in the $\ge 4l+\ET$ channel, we include the following SM processes: 
\begin{center}
$VVV$, $ZZ$, $t \bar t Z$, $t \bar t $, $b \bar b Z$ and $t \bar t t \bar t$,
\end{center}
where $V=W,Z$ are allowed to be off shell. 

Out of the contributions to $VVV$, $WWZ$ is the most dominant contribution to our background and is the one we include in our analysis. We do not simulate $VVVV$ processes, which are expected to be subdominant \cite{Baer09}. 

We expect to be able to efficiently reduce these backgrounds in order to discriminate the signal: \;$ZZ$ production is the dominant source of hadronically quiet 4$l$ events, but without invisible particles in the final states it can be efficiently removed by a cut on missing transverse energy. For  IDMs producing leptons from off-shell $Z$ bosons, the SM backgrounds including on-shell $Z$ can be further discriminated by reconstructing the invariant mass of same-flavor, opposite-sign lepton pairs. 
The $t\bar t Z$ and $t\bar tt\bar t$ backgrounds can also be reduced by vetoing $b$ tagged jets, which should leave most of the IDM signal events. 
For the low background levels in the four-lepton channel, a significant contribution could come from fake leptons. This is difficult to properly take into account in a study based on Monte Carlo simulation, and should be estimated from experimental data. We comment further on this in our discussion of systematic uncertainties in Sec.~\ref{sec:systematics}.

\section{Analysis} \label{sec:5}
\begin{table*}[t]
    \begin{tabular}{ c|c|c|c|c|c}
    Benchmark 	& $\sigma_{pp\rightarrow H^+H^-}$	& $\sigma_{pp\rightarrow H^+A^0}$	& $\sigma_{pp\rightarrow H^-A^0}$ 	& Br$_{H^\pm\rightarrow A^0}$& $\sigma_{4l}$\\ \hline
    IDM-A1    	& 	18.56 	& 54.00	& 29.19	& 0.191	&0.11         	\\ 
     IDM-A2    	&  	13.36 	& 42.65	& 22.68	& 0.293	&0.12   	 	\\    \hline
        IDM-B1    	& 	69.43  	& 123.8	& 70.51 	& 0.071	& 0.095    		\\ 
   IDM-B2    	&  	16.23 	& 36.36	& 19.12  	& 0.008	& 0.003		\\ \hline
   IDM-C1		&	27.63	& 62.44	& 34.05	& 0.013	& 0.008			\\
   IDM-C2		& 	27.20	& 56.25	& 30.47 	& 0.002	& 0.001			\\
    IDM-C3    	& 	14.05  	& 32.71	& 17.06  	& 0.079	& 0.003       		\\ 
   IDM-C4    	&  	 13.77	& 32.71	& 17.05 	& 0.070	& 0.022         		\\ 
      \end{tabular}
    \caption{Cross sections for processes where the interaction is mediated via gauge bosons $Z/\gamma$ or $W$, in units of fb. }
    \label{tab:2}
\end{table*}
\begin{table*}
    \begin{tabular}{@{\extracolsep{\fill}} c|c|c|c| r} 
    Benchmark 	& $\sigma_{gg\rightarrow A^0A^0}$	& $\sigma_{gg\rightarrow H^+H^-}$& $\sigma_{4l}$	&Br($h\rightarrow 2H^0\!+\!2A^0\!+\!H^+H^-$) \\ \hline
  IDM-A1    	&   88.25	& 7.46	 & 0.40  	&$(0 +0.84 + 0) = 0.84\,\%$\\ 
  IDM-A2    	&   4.66	& 138.0 	 & 0.32      	&$(0 +0.09 +3.6) =3.7\,\%$\\ \hline
  IDM-B1    	&   783.4	& 1198 	& 4.1      	&$(5.9+7.7+13) =27\,\%$			\\ 
  IDM-B2    	&   194.6	& 386.7	 & 0.90 	& $(4.3+5.7+15)=25\,\%$			\\ \hline
      IDM-C1	& 2844	& 162 	& 13		& $(32+30+0) =62\,\%$	 \\
      IDM-C2	& 1981	& 39.55 	& 9.0		& $(27.4+19.4+0) =47\,\%$	 \\
 IDM-C3    	&   483.0	& 558.8	 & 2.5	& $(15+16+28) =59\,\%$			\\ 
  IDM-C4    	&   194.5	& 366.1	 & 1.1	& $(4.5+5.7+15) =25\,\%$			\\
    \end{tabular}
    \caption{Cross sections for processes where the interaction is via the Higgs boson, in units of fb.}
    \label{tab:2b}
\end{table*}
In order to study the signal expectations for IDM at the detector level, we define a set of benchmark models in Table~\ref{tab:1}. The models are divided into three subsets:
\begin{itemize}
\item The {\bf A} models are constructed to test how strong the signal can be when inert states are produced via gauge interactions, and the direct detection signal will be very weak. These models do, however, give invisible Higgs branching ratios that are too low to pass the current LHC constraints on a heavy Higgs, and are therefore ruled out but kept here for illustration of the strength of the gauge-mediated production.
\item  The {\bf B} models represent IDM scenarios that explain all of the observed DM. They only pass all constraints if we add the systematic uncertainties to the XENON-100 limits, as discussed in Sec.~\ref{sec:4B}. 

Models with a Higgs mass above $600$ GeV (where no Higgs search limits have been presented) should more easily pass all constraints. Such models should be able to give similar $4l+\ET$ signal features as IDM-B2. However, a weaker signal is expected since the production cross section of inert states will be smaller once $\lambda_{H^0}$ is adjusted to pass the direct detection constraint (unless we again allow for systematic uncertainties as for IDM-B2).

The model IDM-B1 has a Higgs boson mass of 320 GeV, motivated by the excess seen by the CMS experiment around this value. This possibility was, however, ruled out by the new ATLAS limits \cite{atlas-combined-update} that were presented during the preparation of this manuscript (see Figure~\ref{fig:higgsbranch}). 
\item  The {\bf C} models are illustrative examples of models that pass all constraints, but have a relic density that explains only a fraction of the observed total cold DM content. They are chosen such that IDM-C1 and IDM-C3 just pass the XENON-100 constraint, but have some margin to the LHC Higgs bound. IDM-C2 and IDM-C4 instead just evade current Higgs searches at the LHC, but have larger margins to the XENON-100 limits. 

The models IDM-C2, IDM-C3 and IDM-C4 give a relic DM contribution of 10\% to $\Omega_{\rm CDM}$, and IDM-C1 gives 1\% of $\Omega_{\rm CDM}$. 
\end{itemize}
All the benchmark models pass all the other experimental and theoretical constraints listed in Sec.~\ref{sec:2}.\footnote{The high Higgs mass in combination with large couplings actually renders the IDM-C3 model marginally in violation of the, somewhat arbitrary, choice for the tree-level unitarity limit given in Sec.~\ref{sec:2}.} 
In our detector-level study, we take these as our representative IDM models for a tetralepton signature with a heavy SM-like Higgs boson.
In Table~\ref{tab:2} and \ref{tab:2b}, we list the models' properties relevant for the four-lepton signal.

These models may well show up in upcoming data from XENON-100 and LHC. The expected performance of LHC is an integrated luminosity of up to $\sim15$ fb$^{-1}$ collected by end of 2012 with an upgrade to 8 TeV for the rest of this year.  The increase in sensitivity in the SM Higgs searches is about a factor 1.6 due to the integrated luminosity being 3 times larger and a factor about 1.2 due to the increased energy \cite{atlas-higgs-sensitivity}. A factor up to about $\sqrt{2}$ could also come from combining the ATLAS and CMS data. This means that all our benchmark models, except possibly IDM-C3, should be reached by exclusion limits from LHC Higgs searches by the end of 2012. Detection of the Higgs bosons in any of our benchmark models, at the $5\sigma$ level, would however require more integrated luminosity, and such Higgs bosons would most likely not be revealed before the LHC run at 14 TeV.

The cross section sensitivity of XENON-100 will also improve by an order of magnitude by the end of 2012 \cite{Aprile:2009yh,XENON_prospects}. This is enough to start to probe all our benchmark models, except for the IDM-C1 and possibly IDM-C4 model (and of course the IDM-A models). The planned XENON-1T is expected to improve the sensitivity by more than an order of magnitude \cite{Aprile:2009yh,XENON_prospects}. 

We therefore consider the complementary four-lepton plus missing energy channel as a potential step to further pin down or discover an IDM signal. 

\subsection{Event generation} \label{sec:eventgeneration}
We generate signal and background events with the \mg/\me-4.4.32 package. From a user specified process, \mg creates Feynman tree-level amplitudes (including effective operators and using a \helas \cite{Murayama:1992gi} implementation for the helicity amplitude calculations) for all relevant hard subprocesses. Once events are generated with \me they are passed to \pythia \cite{Sjostrand:2006za} for hadronization and decay. The events are then passed to the Pretty Good Simulator \pgs \cite{PGS} to mimic the detector response.

For each background and signal process, we generate events corresponding to an integrated luminosity of at least 10 times the integrated luminosity for which we make predictions. In a few cases, however, we were limited by computer power, and for the IDM-A models and the SM backgrounds we have generated events corresponding to at least 3000 fb$^{-1}$, except for $b\bar bZ$ and $t\bar t$ production for which we have generated 220 fb$^{-1}$ and 160 fb$^{-1}$, respectively.

\subsection{Settings} 
We consider proton-proton collisions at 14 TeV, using the standard \textsf{cteq6l1} for the parton distribution functions \cite{Pumplin:2002vw}. In \pythia, we include initial- and final-state radiation but not multiple interactions. For our \pgs settings we choose the options that mimic the ATLAS detector with a cluster finder cone size of $\Delta R=0.4$ for jet reconstruction, and keep the other parameters as they are given by default in pgs\_card\_ATLAS.dat in \mg/\me-4.4.32.

For the cases where we generate events including jet matching (see Sec.~\ref{sec:systematics}), we use the so-called MLM scheme \cite{MLM,Mangano:2006rw} with the minimum $K_T$ jet measure for the phase space separation between partons set to 20 GeV.

The lepton isolation criteria are an important part of the lepton object definition in order to distinguish them from leptons that could have originated in jets. For electrons, \pgs does this by default by requiring that the transverse calorimeter energy in a ($3\times 3$) cell grid around the electron, excluding the cell with the electron, has to be less than 10\% of the electron's transverse energy and that the summed $p_T$ of tracks within a $\Delta R=0.4$ cone around the electron, excluding the electron, is less than 5 GeV.  To mimic the ATLAS detector response, we also ignore electrons with a pseudorapidity $\eta$ within $1.37\le|\eta|\le1.52$ \cite{Aad:2009wy}. For muons, that are not isolated by default in \pgs (and we do not make use of the cleaning script that is the default in \mg/\me), we require the summed $p_T$ in a $\Delta R=0.4$ cone around them, excluding the muon itself, to be less than 10 GeV to define them as isolated. For each lepton we also require a minimum distance of $\Delta R=0.4$  from the nearest lepton or jet (as reconstructed by PGS).


\begin{figure*}\
\centering
\begin{tabular}{cc}

\begin{overpic}[height=7cm, width=8cm]{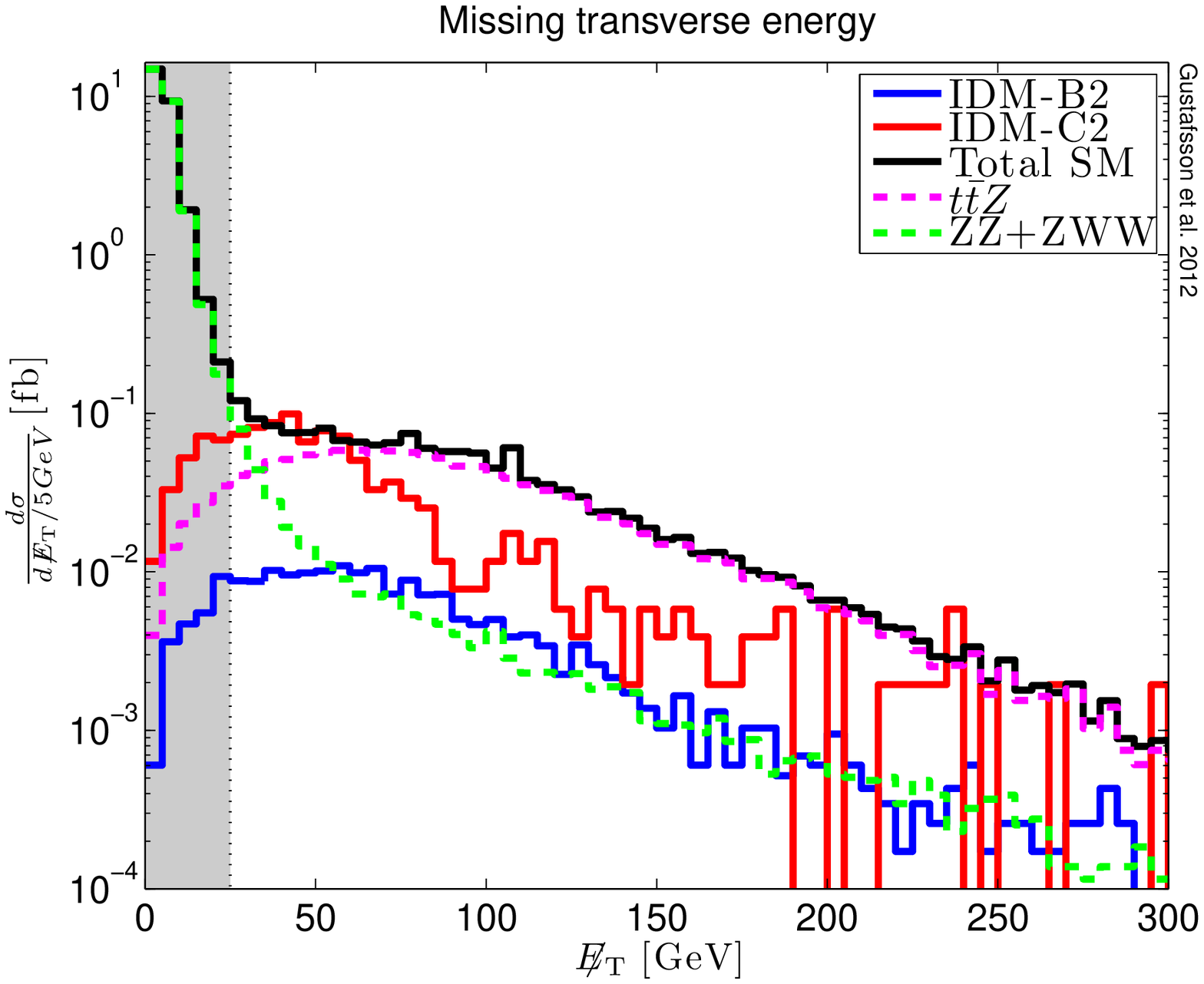}
\put(55,139){\scriptsize Total SM} 
\put(52,141){\color{black}\vector(-1,-0){12}}
\put(75,135){\color{black}\vector(1,-2){12}}
\put(39,68){\scriptsize IDM-B2} 
\put(105,80){\scriptsize IDM-C2} 
\put(55,100){\scriptsize $t\bar{t}Z$} 
\put(40,89){\scriptsize $ZZ\!+\!ZWW$} 
\end{overpic}
\begin{overpic}[height=7cm, width=8cm]{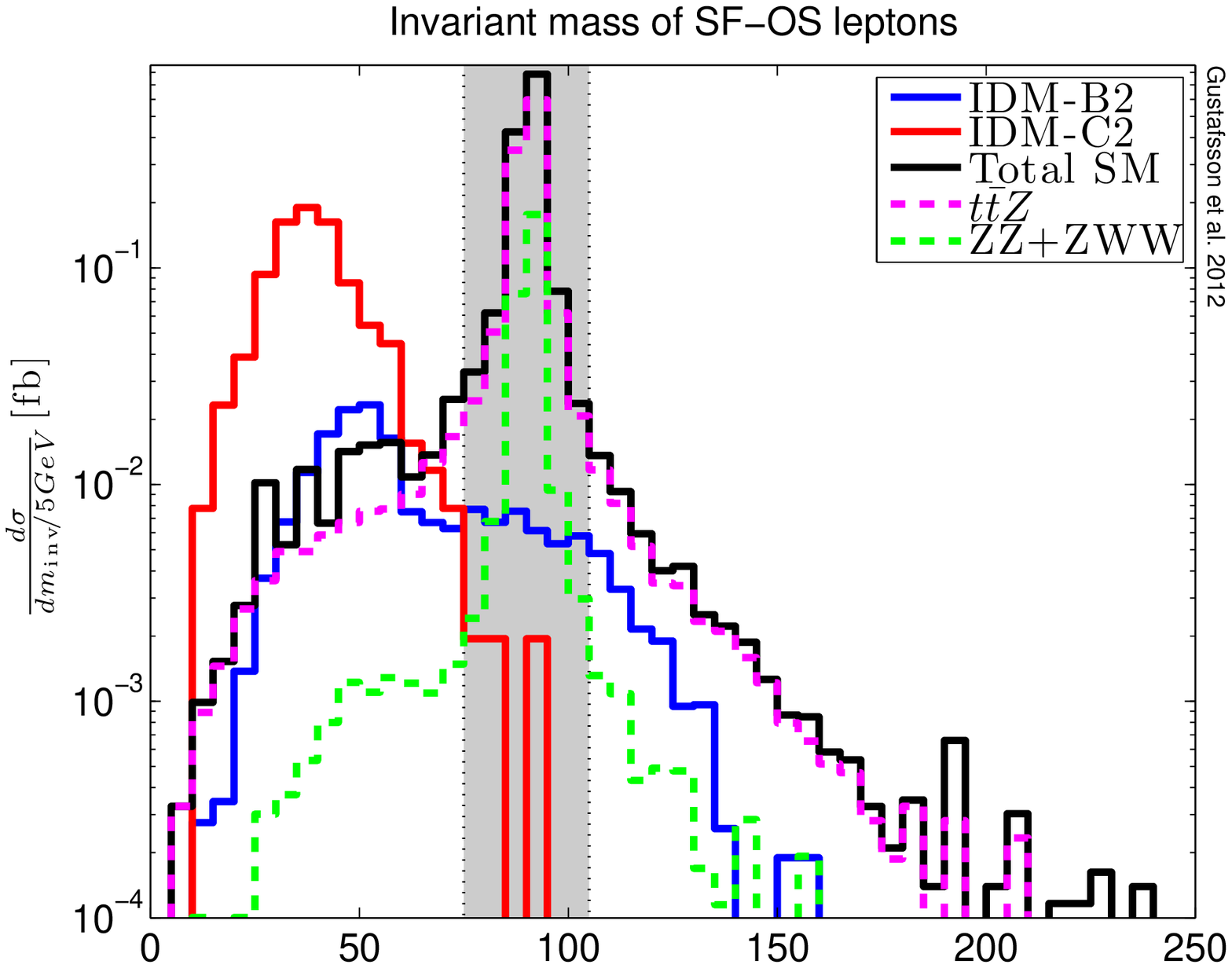}
\put(105,175){\scriptsize Total SM} 
\put(45,117){\scriptsize IDM-B2} 
\put(42,156){\scriptsize IDM-C2} 
\put(60,80){\scriptsize $t\bar{t}Z$} 
\put(45,45){\scriptsize $ZZ\!+\!ZWW$} 
\end{overpic}
\\
\begin{overpic}[height=7cm, width=8cm]{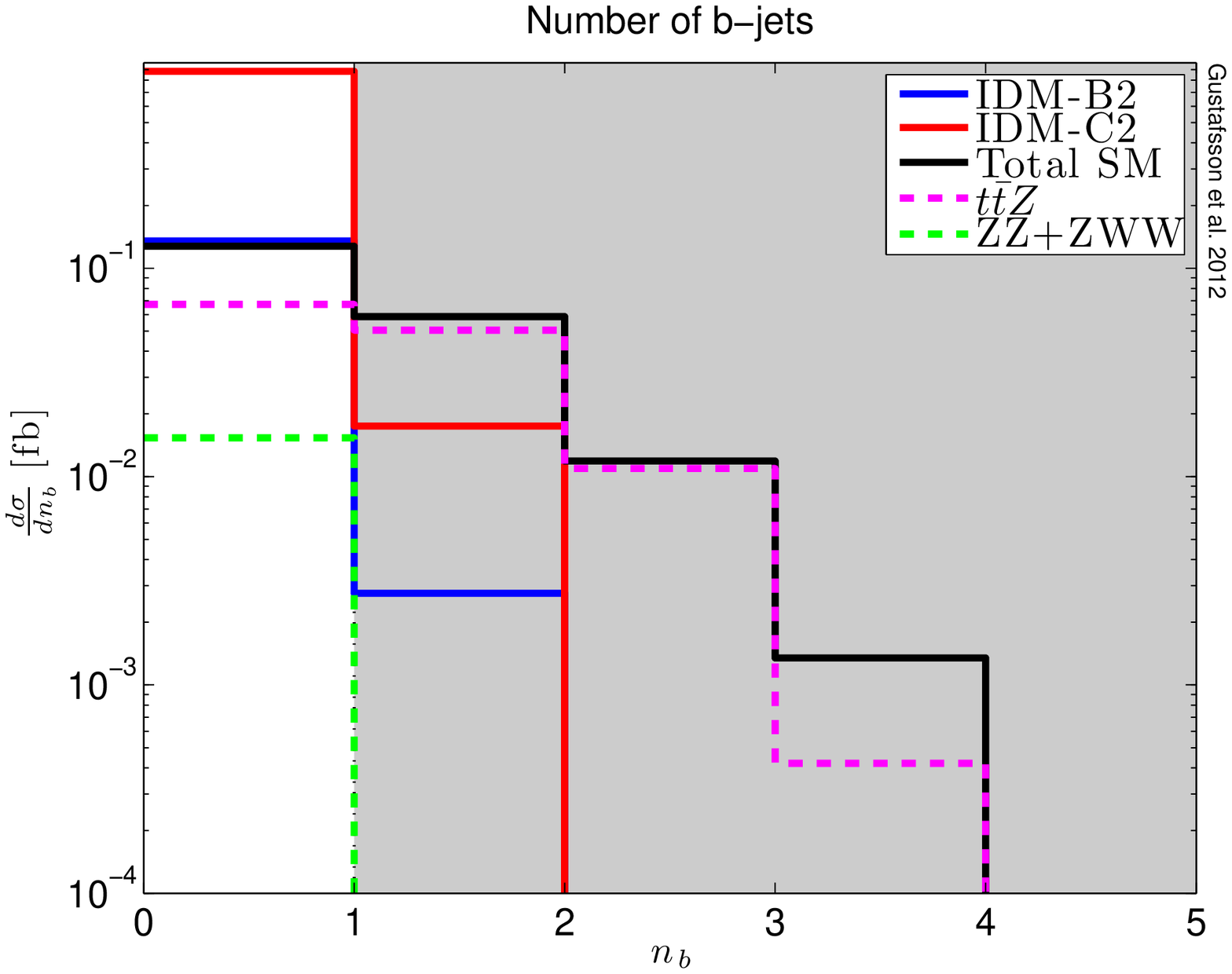}
\put(95,145){\scriptsize Total SM} 
\put(90,149){\color{black}\vector(-1,0){22}}
\put(90,145){\color{black}\vector(-1,-1){10}}
\put(125,143){\color{black}\vector(0,-1){37}}
\put(70,80){\scriptsize IDM-B2} 
\put(30,150){\scriptsize IDM-B2} 
\put(68,170){\scriptsize IDM-C2} 
\put(35,125){\scriptsize $t\bar{t}Z$} 
\put(30,100){\scriptsize $ZZ\!+\!ZWW$} 
\end{overpic}
\begin{overpic}[height=7cm, width=8cm]{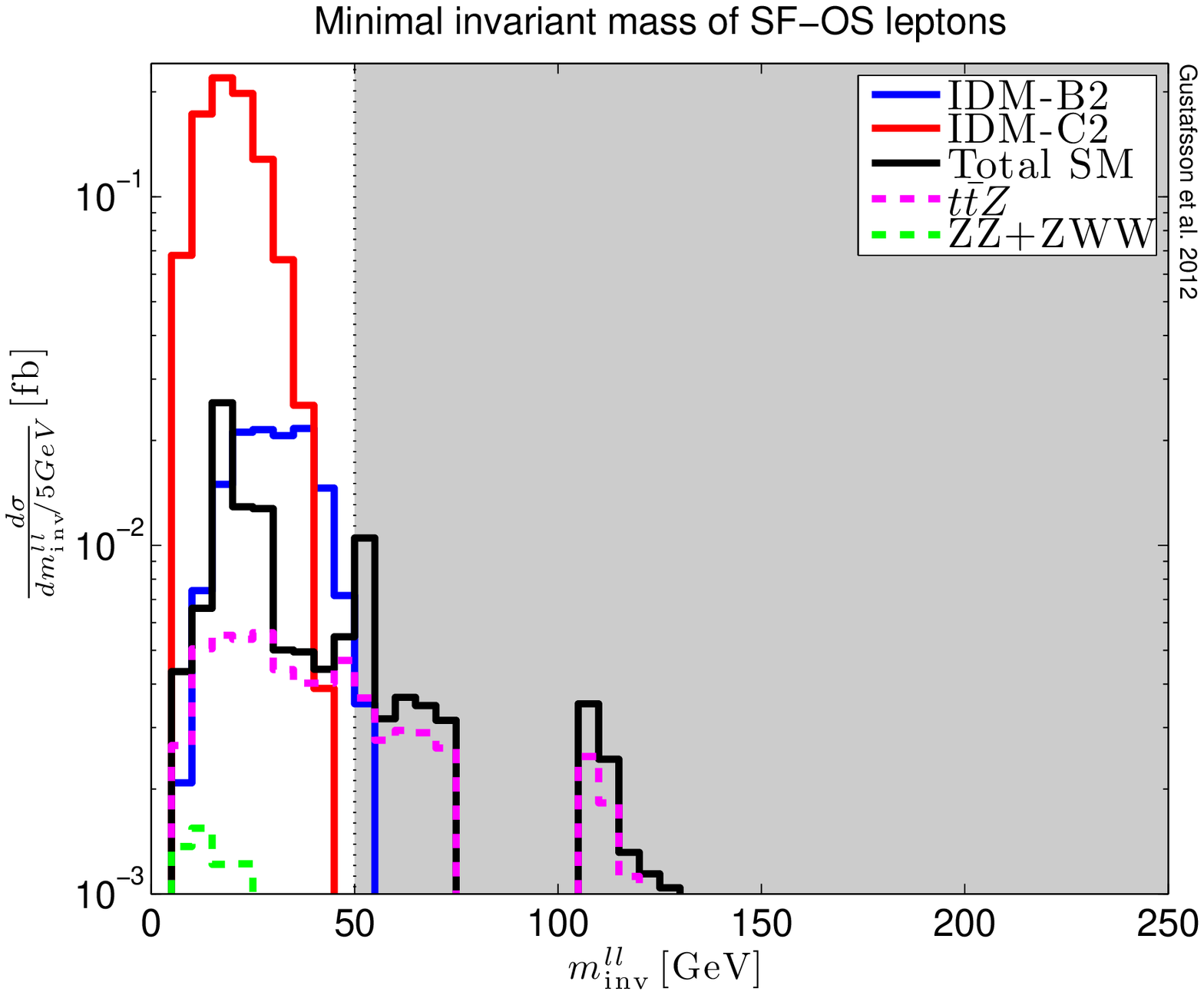}
\put(40,120){\scriptsize Total SM} 
\put(60,110){\scriptsize IDM-B2} 
\put(50,170){\scriptsize IDM-C2} 
\put(40,60){\scriptsize $t\bar{t}Z$} 
\put(40,30){\scriptsize $ZZ\!+\!ZWW$} 
\end{overpic}
\end{tabular}
\caption{Top left: Missing transverse energy distributions in events with four isolated leptons. Top right: The invariant mass of SF-OS lepton pairs after the cut on $\ET$ has been applied. Bottom left: Distribution of $b$ tagged jets, after the cut on $\ET$ and $Z$ veto. Bottom right: Invariant mass distribution for the SF-OS lepton pair producing the minimal such value per event, after all other cuts have been performed. The shaded grey regions indicate the cuts on each quantity.
}
\label{fig:distributions}
\end{figure*}


\subsection{Cuts} \label{cutsection}
In order discriminate an IDM signal from SM background events, we perform cuts sequentially on the detector simulator's reconstructed particle data. 

To illustrate our cuts, we show in Figure~\ref{fig:distributions} the event distributions after each cut. The plots include two of our benchmark models together with the total SM background and two of its main subprocess contributions in this tetralepton + $\ET$ channel. These are the cuts specific to our IDM study:
\begin{itemize}
\item 
First, we require four or more isolated leptons. In order to make lepton isolation and event triggering in the four-lepton channel robust, we will require a leading lepton with $p_T^{l_1}\ge20$ GeV and that each of the additional leptons have $p_T^{l_{2,3,4}}\ge$ 10 GeV. 
\item
In order to reduce the $ZZ$ background efficiently, we require the missing transverse energy ($\ET$) in each event to be larger than 25 GeV, as illustrated in the (upper left) panel of Figure~\ref{fig:distributions}.  
\item
We reject events with \emph{any} pair of same flavor and opposite sign (SF-OS) leptons among the $\geq4$ leptons with an invariant mass that falls within the range of the $Z$ resonance, 75~GeV$<m_{\rm inv}^{l^+l^-}<105$~GeV. We refer to this as our $Z$ veto. The (upper right) panel in Figure~\ref{fig:distributions} shows the distribution of events by the pair of SF-OS leptons giving an invariant mass closest to 91 GeV.
\item
The $t\bar t Z$ background can be fairly efficiently discriminated against by requiring no $b$ tagged jets in the event, as illustrated in the (bottom left) panel of Figure~\ref{fig:distributions}. Because of the displaced vertices from $b$ quarks, the background from $t\bar t$ and $b\bar b Z$ could be further discriminated against by using a cut on the impact parameter for muons \cite{b-leptons}. Such an improvement is beyond the scope of this paper, since it cannot be done within the standard \pgs detector simulation that we use.
\item
In the (bottom right) panel of Figure~\ref{fig:distributions}, we show the distribution of events in the minimal SF-OS dilepton invariant mass (minimal since each event has at least four leptons, and may contain more than one pair of SF-OS leptons). This invariant mass is expected to be low for our benchmark models, as the $Z$ decays off shell, and we require the minimal invariant mass to be $<50$ GeV. 
\end{itemize}
For the signal events, the position of the peak in the SF-OS dilepton invariant mass distributions is slightly below the mass difference $\Delta m_{A^0H^0}$ in a given model. The large fluctuations in the minimal invariant mass distribution of the total SM background (bottom right panel of Figure~\ref{fig:distributions}) come from the low statistics of our $t\bar t$ sample; only six $t\bar t$ events are left after the cuts, five of which lie in the 15-25 GeV bins. This makes it difficult to say something about the distribution of this specific background contribution. What we can see is that if the $t\bar t$ events could be vetoed in some way, for example using the impact parameter for muons mentioned above, then the SF-OS dilepton invariant mass distribution can be used as a signature to clearly distinguish our models from the background.

A characteristic of our benchmark models is that the signal leptons originate in off-shell $Z$ bosons. Therefore our signal efficiency is sensitive to the isolation criteria and the minimum $p_T$ requirements on the leptons. Models with larger $\Delta m_{A^0H^0}$, which allow $A^0$ to decay to on-shell $Z$, would be more difficult to detect since in this case  the signal cannot be distinguished from the background using the $Z$ veto.

\subsection{Sources of systematic uncertainties}\label{sec:systematics}

Systematic uncertainties in our $4l+\ET$ signal study are due to limited statistics in some of our background samples, sensitivity to lepton efficiencies and fake-lepton contributions to the background.

\begin{table}[!h]
    \begin{tabular}{@{\extracolsep{\fill}} c|c|c|c|c|c}
    Process 	&$n_l\ge4$	&$\ET$ cut	& $Z$ veto&$n_b=0$&$m_{\rm min}^{l+l-}$ cut	 \\ \hline  
    $WWW$	& 0.0049		& 0.0025		& 0.0025	& 0.0025	& 0 ($<0.0025)$					 \\ 
    $WZ(j)$		& 3.4     		& 2.2			& 0.24	& 0.24	& 0 ($<0.059)$				\\ 
$ZZ(j)$    		& 2900		& 23   		& 0.59 	& 0.53	& 0.46				\\ 
    $t\bar tW(j)$	&  1.1		& 1.1			& 0.80	& 0.47 	& 0.19	 				\\ 
    $t\bar tZ(j)$	& 150		& 140		& 13		& 6.3		& 3.5					\\ 
   $t\bar tt\bar t(j)$& 0.62		& 0.61		& 0.41      	& 0.14	& 0.038					\\ 
      \end{tabular}
\caption{Cross sections for backgrounds that require fake leptons to produce four final-state leptons (the WWW,WZ processes). We also test if our backgrounds are sensitive to including jet matching. None of these effects seem important if compared to the values used in our final analysis presented in Table~\ref{tab:3new} (where jet matching is not included). We have required four isolated leptons and add the respective cuts for each column successively. Results are presented in units of $10^{-2}$ fb. We have included $K$ factors of 1.6 for $ZZ(j)$, 1.9 for the $WZ(j)$ \cite{Campbell:2011bn} and 1.4 for $t\bar{t}Z$ \cite{Lazopoulos:2008de}.}\label{tab:5}
\end{table}

In our statistical analysis we fix the signal and background cross section expectations to our average results but, as mentioned in Sec.~\ref{sec:eventgeneration},  generation of enough $b\bar b Z$ and $t\bar t$ events were limited by computer power. We trust that our cuts remove any contribution from $b\bar b Z$, but the $t\bar t$ contribution gives an uncertainty in our background estimation. Our $t\bar t$ sample consists of only five events after all cuts, and for a Poisson distribution the upper expectation value is 9.3 events at 90\% CL Taking this upper value as the average $t\bar t$ result instead would increase our total background cross section with only 30\% (and a similar relative increase in the expected needed luminosities to discover the signals).

The lepton efficiency is low for our models compared to the SM background. This is because the leptons in the model events originate in off-shell $Z$ boson decay, and our signal predictions are thus sensitive to the lepton isolation and $p_T$ requirements. For comparison, if we use our $p_T^l$ requirements and decrease the lepton efficiencies, as in \cite{Edsjo:2009rr}, both the signal  and the SM background cross sections are reduced by about 50\%. Because of the increase of pile-up effects as the experiment reaches design luminosity, the isolation criteria might have to be loosened and the $p_T$ requirement raised in compensation. For our study, a raise to $p_T^{l_{2,3,4}}>15$ GeV would  leave only 32\% of the total background (completely remove the contribution from $t\bar t$), while still leaving 20\%-60\% of the signal in our benchmark models.

Since the background in this channel is low, we could be very sensitive to the contribution from fake leptons. In order to make use of PGS's ability to generate fake electrons, we show in Table~\ref{tab:5} the results of our cuts applied on some SM processes that naively give three lepton final states (such as $WZ$) and include explicit jets. 
The table shows that these types of fake-lepton contributions seem not to be very important. Likewise, we find that including jet matching would not alter the result in our final analysis that is presented in Table~\ref{tab:3new} (where jet matching, for consistency, is not included for neither the backgrounds nor the models).

A proper inclusion of backgrounds involving fake leptons has to be based on experimental data. In a recent ATLAS analysis \cite{atl-4l-jan12} of the $4l+\ET$ channel, the systematic uncertainty due to differences in fake rate between simulation and data was estimated to be around 10\% for  the background processes $t\bar t$ and $t\bar t Z$. They also find that the $Z+$jets give a significant contribution to the background, potentially dominated by electron Bremsstrahlung in the detector material that subsequently pair produce leptons. However, these events are found to contain $\ET$ of 20-60 GeV and hard jets, as can be seen in Figure~2 in \cite{atl-4l-jan12}.  Requiring $<3$ jets with $p_T>40$ GeV and optimizing the $\ET$ cut could potentially reject this background effectively without loss of more than $\sim$10 \% of the signal events in our benchmark models. The uncertainties in the estimation of the $Z+$jets contribution are however large and an inclusion of this background is beyond the scope of our phenomenological study.

Sources of systematic uncertainties will not be included in the following statistical analysis.

\subsection{Results}\label{sec:6}

\begin{table}
    \begin{center}
    \begin{tabular}{@{\extracolsep{\fill}} c|c|c|c|c|c}
    Proc./Model 	&$n_l\ge4$	&$\ET$ cut	& $Z$ veto	& $n_{b}=0$		&$m_{\rm min}^{l+l-}$ cut					\\ \hline
    $ZWW$		&	15		& 13			& 0.92		& 0.92			& 0.42	 				\\ 
      $ZZ$    	&	2700		& 16	 		& 0.62		& 0.62			& 0.47					\\ 
  $t\bar tZ$	&	130		& 120 		& 13 			& 6.7	 			& 4.1					\\ 
  $b\bar bZ$	&	7.2		& 0.89		& $0(<0.45)$	& 0				& 0								\\
   $t\bar t$ 		&	7.6		& 6.9			& 5.0			& 4.4				& 3.2								\\
   $t\bar tt\bar t$	&  	0.56		& 0.56		& 0.46		& 0.093			& 0.031					\\ \hline
 Total bkg		&  	2900		& 160	 	& 20 			& 13				& 8.2					\\ \hline
    IDM-A1		& 	4.6		& 3.5			& 3.3			& 3.2				& 3.2					\\
    IDM-A2    	& 	7.8		& 7.1			& 5.7			& 5.5				& 5.5 				\\ \hline
       IDM-B1	& 	17		& 14			& 13			& 13 				&  13	 				\\
    IDM-B2		& 	20		& 18			& 14			& 14				& 13						\\ \hline
    IDM-C1		&	41		& 31			& 29			& 28				& 27				\\
    IDM-C2		&	110		& 90			& 90			& 88 				& 88				\\
       IDM-C3	& 	34		& 30			& 26			& 26				& 26					\\
    IDM-C4		& 	22		& 19			& 15			& 15				& 15	 			\\ 
      \end{tabular}
    \end{center}
\caption{Results of the SM background and total IDM signal cross sections in units of $10^{-2}$ fb. We have required four isolated leptons and for each column, from left to right, we successively add our other cuts as described in the text. $K$ factors of 1.6 for the $ZZ$ background \cite{Campbell:2011bn} and 1.4 for $t\bar{t}Z$ and $t\bar{t}$ \cite{Lazopoulos:2008de} have been applied, as well as the $K$ factors for the signal processes as previously quoted.}
    \label{tab:3new}
\end{table}

\begin{table*}[]
    \begin{center}
    \begin{tabular}{@{\extracolsep{\fill}} c|c|c|c|c|c|c|c|c}
 Model										&IDM-A1	&IDM-A2	&IDM-B1	&IDM-B2	&IDM-C1	&IDM-C2	&IDM-C3	&IDM-C4		\\ \hline
 $3\sigma$ evidence, $P_\text{obs}=50\%$ (fb$^{-1}$)	& 810	& 300	& 64		& 64		& 19		& 3.8		& 20		& 50			\\ 
 $5\sigma$ detection, $P_\text{obs}=50\%$ (fb$^{-1}$)	& 2300	& 820	& 180	& 180	& 53		& 9.0		& 55 		& 140				\\ 
 95\% CL exclusion, $P_\text{obs}=50\%$ (fb$^{-1}$)	& 280	& 110	& 30		& 30		& 13		& 3.1		& 14		& 20				\\ 
      \end{tabular}
    \end{center}
\caption{The expected integrated luminosities needed at 14 TeV for a 3$\sigma$ and 5$\sigma$ detection in the inert doublet benchmark models. Alternatively, the expected luminosity needed for a 95\% CL exclusion of these benchmark models.}
    \label{tab:4}
\end{table*}

In Table~\ref{tab:3new}, we show the results after the signal and background events have been passed through the \pgs detector simulation as we successively perform  the cuts described in Sec.~\ref{cutsection}. 

To obtain a statistical measure for when our signal could be observed or excluded, we assume the number of events to be Poisson distributed. The probability of observing $N$ or fewer events is then
\beq
P(N;B)=\sum_{n=0}^{N}\frac{B^ne^{-B}}{n!},
\eeq 
given that the background expectation value $B$ is the true mean. For a one-sided 3(5)$\sigma$ detection, we take the probability $(1-P(N;B))$ of having this number of events or more due to a statistical background fluctuation to be less than 0.13\,\% ($2.9\times10^{-5}$\,\%). With a signal expectation $S$, the probability to observe such an excess signal is \mbox{$1-P(N;S+B)$}, and we request this probability $P_{obs}$ to be 50\%.

In Table~\ref{tab:4} we show the prospects for when detection or exclusion of our benchmark models at the LHC will occur. The quoted integrated luminosities are for a 50\% probability to have at least $3\sigma$ evidence, $5\sigma$ detection or 95\% CL exclusion of the models. 

Because of the sometimes low statistics needed to detect these models, the use of Poisson statistics should be more correct than {\it e.g.}\ the commonly used rule of thumb of a 5$\sigma$ discovery when $S > \mathrm{max}(5, 5 \sqrt{B})$. In Appendix \ref{appendix:B} this and other commonly used statistical measures are compared. For the benchmark models with the strongest signal, and thus the lowest number of expected events at the time of a discovery, Poisson statistics lead to about a factor of two larger required integrated luminosity than a naive Gaussian approximation. In Appendix \ref{appendix:B}, we also show that increasing the prospect from 50\% to 90\% probability to find evidence for a signal can require up to a factor of two increase in the required integrated luminosity.

\vspace{-0.2cm}
\section{Conclusions}\label{sec:7}
We have investigated the status of the IDM in light of the results from DM direct detection searches by XENON and the Higgs searches at the LHC. These experimental results complement each other in constraining the viable parameter space. 

We first set out to study the IDM in the regime where the model both provides a DM candidate and includes a heavy Higgs boson, thereby possibly alleviating the LEP paradox. Considering the model's ability to evade the SM Higgs searches, we investigated the effect of imposing the bounds from direct detection assuming that the DM abundance is set by thermal freeze-out of the inert $H^0$ particle. 
 
In particular, the combination of constraints utilized in this work completely rules out the so-called `new viable region' found in \cite{LopezHonorez:2010tb} where $H^0$ masses are in the range of 80-150 GeV. Moreover, we conclude that the ensemble of constraints are in conflict with the IDM for its whole viable cold DM mass range \emph{if} the models shall also incorporate the Higgs boson in the mass range 160-600 GeV. This conclusion can be avoided if either 1) the canonical experimental bounds can be relaxed or 2) the IDM does not account for all the DM. 
  
We investigate the prospects of detection/exclusion in the near future of models belonging to these types of 'escape' scenarios.
Adding the systematic uncertainties to the observational constraints, and at the price of some fine-tuning,  we found that we can still obtain IDMs that contain both a heavy Higgs boson ($\gtrsim$500 GeV) and a good DM candidate. We also looked into the possibility that IDM explains only a fraction of the universe's DM content, and thereby more easily evades current constraints from both LHC and DM direct detection experiments. Some of these models can be efficiently probed by the foreseen data from XENON and LHC before the end of 2012. 

\medskip
The potential detection of a heavy Higgs boson and/or a signal in direct DM detection experiments in the viable IDM DM mass range, although these would be striking features in favor of an IDM-like scenario, would not exclusively point to the IDM. A way to pin down the identity of the new physics further would be to compare different complementary channels. The prospects for detection of the IDM in the 14 TeV LHC data have been studied previously for channels with two or three leptons, together with missing energy  \cite{Dolle:2009ft, Miao:2010rg}. In this work, we have investigated the possibility of a four-lepton plus missing energy signature at the LHC coming from IDM. The models with a heavy Higgs boson that evade the current constraints typically have large couplings between the inert states and the SM-like Higgs boson. As a result, the production of four-lepton final states via gluon fusion Higgs production becomes a particularly promising channel to track, and even discover, the IDM during the early runs at LHC's design center-of-mass collision energy. 

We find that in the four-lepton plus missing energy channel our benchmark points, where the inert particles are mainly produced via the Higgs boson, should show up early in the 14 TeV LHC run. 
Our models IDM-B1, IDM-B2 and \mbox{IDM-C1 to IDM-C4)} should be seen at integrated luminosities of 3.8-64 fb$^{-1}$ (9-180 fb$^{-1}$) at the $3\sigma$ ($5\sigma$) CL. We can note that the IDM benchmark points that were studied in the previous works \cite{Dolle:2009ft, Miao:2010rg} for the dilepton and trilepton channels, only one survives the current direct DM detection and SM Higgs searches. Nevertheless, according to these references, our benchmark points satisfy properties, such as favorable $\Delta m_{A^0H^0}$, that should also render them detectable in the dilepton and trilepton channels at integrated luminosities of 100-300 fb$^{-1}$. We thus conclude that, compiling recent experimental constraints, the IDM with a SM-like Higgs boson heavier than about 160 GeV could very well first show up in the tetralepton channel.

\bigskip
\newpage
\medskip
\noindent
{\bf Acknowledgments:}
The authors thank S.\ Andreas, A.\ Bharucha, E.\ Bergeaas Kuutmann, F.\ Bonnet, Q.\ Cao, J.\ Edsj\"{o}, J.\ R.\ Espinosa, J.\ Sj\"{o}lin, S. Strandberg, R.\ Sundberg, K.\ Tackman  and M.\ Tytgat for discussions. M.G.\ also thanks M.\ Goebel for providing the SM prediction of the oblique parameters from \gfitter. M.G.\ thanks the Fondazione Cariparo Excellence Grant `LHC and Cosmology', the Belgian Science Policy (IAP VI/11: Fundamental Interactions), the IISN and the ARC project `Beyond Einstein: fundamental aspects of gravitational interactions', and S.R.\ the Swedish Research Council (VR) for financial support. L.L.H.  acknowledges partial support from the European Union FP7 ITN INVISIBLES (Marie Curie Actions, PITN- GA-2011- 289442).

\begin{widetext}
\section*{NOTE ADDED AFTER PUBLICATION}

Given the latest results from the LHC of a discovery of a resonance at $\sim126$ GeV consistent with the SM Higgs boson \cite{:2012gk,:2012gu}, we present in Tab.~\ref{tab:125} five IDM benchmark models representative of a SM-like Higgs mass around 126 GeV \cite{CMS:1TeV}.
We have also checked that these are in agreement with the latest results of the XENON-100 experiment \cite{Aprile:2012nq}. A more precise loop calculation in \cite{Klasen:2013btp} typically renders $\sigma_{\text SI} \gtrsim 10^{-47}$cm$^2$, and should shift a few of our lowest  $\sigma_{\text SI}$ to values close to that bound. The viable DM mass region is $50 \lesssim m_{H_0}\lesssim75$ GeV with the updated constraints.

The first two models give a non-negligible contribution to the Higgs width through $h\rightarrow H_0H_0$. None of the other models give a significant contribution to the invisible Higgs width. The third model is indeed chosen to have a negligible coupling to the Higgs, although the $H^0$ mass is slightly below half the mass of the Higgs, and the next three models have inert masses that are too large for the particles to be produced on the Higgs resonance. For the fourth model, the relic density is obtained through coannihilations resulting in a small $\sigma^{\text{SI}}$. For the fifth point, the annihilation cross section receives non-negligible contributions from both 2- and 3-body final states. The generation of the relic density for the last point is entirely driven through annihilations into  $WW$,  2-body processes 
, although the zero velocity annihilation cross section is driven by 3-body processes. The first and the last benchmark points lie on the limit of the viable $H_0$ mass range.

Note that these types of models were not explicitly included in our four-leptons analysis. Inert scalars produced via a 126 GeV SM-like Higgs boson can only decay to $A_0$ particles with masses below $m_h/2$. Therefore, unless the $H_0$ mass is below $\sim 40$ GeV, the final state leptons from $A_0$ decays into $H_0$ will be too soft to be detected as isolated leptons. However, an $H^0$ mass below at 40 GeV are  too low to give the correct dark matter abundance and to be compatible with our other experimental constraints. Therefore these models never give a strong tetralepton signal. 
In the case in which inert particles are produced via the gauge bosons, the four lepton production is independent of the Higgs mass. The production is governed by electroweak couplings, and there is no resonance like in the case of production via the Higgs boson. We have already argued in Sec.~V. A. that in such a scenario, the detection of a tetralepton signal at the LHC is unfeasible. Such processes could however eventually give rise to di- and tri lepton signals \cite{Dolle:2009ft,Miao:2010rg}. Our models are chosen in such a way as to be similar to the benchmark models in \cite{Dolle:2009ft,Miao:2010rg} while still being consistent with all the most recent constraints included in our work. See also \cite{Goudelis:2013uca} for three proposed benchmark models.

\begin{center}
\begin{table}[h!]
    \begin{tabular}{cccccccccccccc} 
\hline
\hline
  $m_{H^0}$	& $m_{A^0}$	&$m_{H^\pm}$	& $\mu_2^2$
& $\;\;\;\lambda_{H^0}$\;\;\; & $\lambda_{A^0}$ & $\lambda_{H^\pm}$ &$\sigma v_\mathrm{tot / 3-body}$&$\sigma v_\mathrm{\gamma\gamma / \gamma Z}$&$\sigma^{\text{SI}}$&$\Omega_{H^0} h^2$&Br($h\rightarrow$inv)	\\ 
\hline
53.0& 120&130&$2100$&$0.023$&0.39&0.47&0.097/0.0089&1.8/0.095&1.5&0.115&26\%\\
54.0& 140&110&$2500$&$0.013$&0.54&0.31&0.056/0.016&2.3/0.17&0.50&0.107&11\%\\
60.0& 160&160&$3624$&$-7.6\cdot10^{-4}$&0.70&0.70&0.16/0.15&4.5/0.14&0.0015&0.110&0.02\%\\
65.0&72.9&120&$4200$&$8.0\cdot10^{-4}$&0.036&0.32&0.40/0.38&5.7/3.1&0.0013&0.109&0\\
65.0&120& 150&$3640$&$0.019$&0.34&0.60&3.1/1.9&20/14&0.69&0.110&0\\
75.5&130&98.0&$6900$&$-0.038$&0.32&0.086&1.0/0.91&4.5/3.8&2.1&0.104&0\\
\hline
\hline 
      \end{tabular}
    \caption{Benchmark models with $m_h$=126.0 GeV. Masses are given in units of GeV.  Annihilation cross sections, at relative impact velocity $v\rightarrow 10^{-3} c$, are in units of $10^{-26}$ cm$^3$/s for $\sigma v_\mathrm{tot,3-body}$ and in units of $10^{-29}$ cm$^3$/s for $\sigma v_\mathrm{\gamma\gamma,\gamma Z}$. Spin-independent cross sections $\sigma^{\text{SI}}$ are in units of $10^{-45}$ cm$^2$. $\lambda_2 =0.01$ for all models.}
    \label{tab:125}
\end{table}
\end{center}

\end{widetext}

\pagebreak
\newpage

\appendix

\section{Naturalness}\label{sec:naturalness}
The IDM serves as an explicit framework where a heavy Higgs boson, up to around 700 GeV, can be incorporated and still be in agreement with EWPT. While this possibility is interesting in itself, it has also served as an additional motivation for the model. Indeed, a larger Higgs boson mass could alleviate the fine-tuning in the SM and make the model more natural by pushing the need for new divergence-canceling physics to higher energy scales \cite{Barbieri:2006dq}. 

Raising the Higgs mass within the IDM does however not necessarily lead to improved naturalness as compared to the SM \cite{Casas:2006bd}. The new inert scalars contribute with additional corrections to the SM-like Higgs mass, as well as exhibit quadratic divergences of their own. This can lead to increased overall fine-tuning although a larger Higgs mass naively renders it less sensitive to corrections from new physics at high energies. 

\medskip
Let $F^2(p_i)$ be a quantity that depends on some independent input parameters $p_i$. The amount of fine-tuning in $F^2$ associated with $p_i$  can then be taken to be $\Delta_{p_i}^F$, defined by \cite{Barbieri:1987fn}
\beq
\frac{\delta F^2}{F^2}\equiv\Delta_{p_i}^{F}\frac{\delta p_i}{p_i}
\label{eq:Delta_F}.
\eeq
A model is said to be natural, up to an energy scale $\Lambda$, if the total amount of fine-tuning is sufficiently small. The exact upper limit on $\Delta_{p_i}^{F}$ in order for the quantity not to be considered to be fine-tuned is somewhat arbitrary. 

\medskip
The scalar masses are the parameters that receive the dangerous quadratically ultraviolet-divergent contributions. Using momentum cutoff regularization, the one-loop corrections to the scalar mass parameters $\mu_i^2 = {\hat\mu_i^2} +\delta \mu_i^2$ can be written (as in \cite{Casas:2006bd})
\bea
\delta \mu_1^2=\frac{3}{64\pi^2}\left[-8\lambda_t^2\Lambda_{t}^2+(3g^2+g'^2)\Lambda_{1g}^2+8\lambda_1\Lambda_{11}^2\right.\nonumber\\
\left.+\frac{4}{3}(2\lambda_3+\lambda_4)\Lambda_{12}^2\right]
\eea
and
\bea
\delta \mu_2^2=\frac{3}{64\pi^2}\left[(3g^2+g'^2){\Lambda}_{2g}^2+8\lambda_2{\Lambda}_{22}^2\hspace{1.3 cm} \mbox{ }\right.\nonumber\\
\left.+\frac{4}{3}(2\lambda_3+\lambda_4){\Lambda}_{21}^2\right]
\eea
where $\lambda_t$ is the top Yukawa coupling, $g'$ and $g$ are the U(1) and SU(2) gauge couplings and we have assumed independent cutoffs $\Lambda_i$. The loop contribution from internal gauge fields are sufficiently small that $\Lambda_{1g}$ will be irrelevant compared to $\Lambda_{t}$. For large scalar couplings the most relevant ones will be $\Lambda_{11,12}$ and $\Lambda_{22,21}$ -- the momentum cutoffs of the loop contributions from fields associated with the SM doublet and the inert doublet, respectively. In our case the relevant fundamental parameters are $\Lambda_i^2,\lambda_i \in p_i $. We will start by focusing on the $\Lambda_i$ to assess the modelÕs sensitivity to physics at higher energy scales.

\begin{figure}[t] 
\includegraphics[width=0.99 \columnwidth]{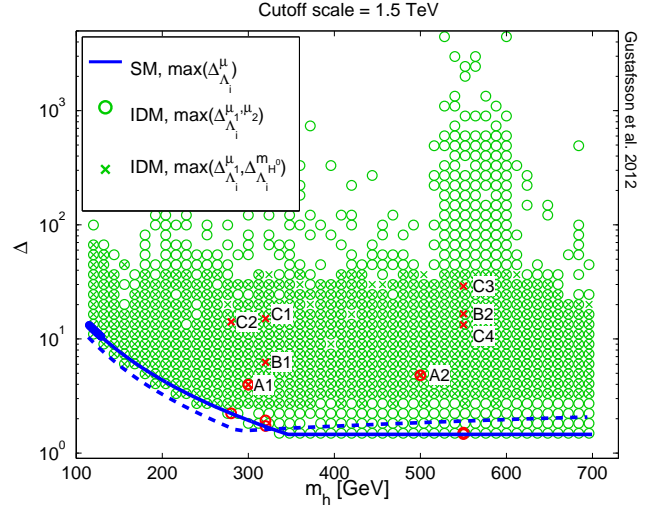}
 \caption{Fine-tuning $\Delta$, without RG effects, as a function of the SM Higgs mass. IDMs are represented by dark (red) marks for the benchmark models and light (green) ones for models in the scan. The SM, given a cutoff scale of 1.5 TeV, is represented by the (blue) solid line. The circles show the result for IDMs using Eq.~\ref{eq:Delta_mu} and the crosses the result using Eq.~\ref{eq:Delta_mH0}. The thick part of the solid (blue) line corresponds to the remaining Higgs mass window allowed within the SM. The dashed (blue) line is the SM result with RG running of the couplings included.} 
 \label{fig:tuning-scan} 
\end{figure} 

\medskip
Taking $p_i=\Lambda_i^2$ for $F^2=\mu_1^2, \mu_2^2$, Eq.~\ref{eq:Delta_F} implies
\beq
\Delta_{\Lambda_i^2}^{\mu_{1,2}}\equiv\frac{\partial{\ln \mu_{1,2}^2}}{\partial{\ln{\Lambda_i^2}}}
\label{eq:Delta_mu}
\eeq
For each model, we take the fine-tuning to be \mbox{$\Delta=\mbox{max}(|\Delta_{\Lambda_i^2}^{\mu_{1,2}}|)$}. Specifying an acceptable level of fine-tuning thus determines the cutoff scale up to which the theory is natural without introducing any new physics. 

In Figure~\ref{fig:tuning-scan} we plot the fine-tuning $\Delta$ for a given cutoff scale of $\Lambda_i=1.5$~TeV. 
This cutoff scale corresponds to the perturbativity scale (in the SM) at which the one-loop RG corrections to the Higgs self-coupling grow to the same level as its tree-level value for a $m_h\sim 700$ GeV. In \cite{Barbieri:2006dq} it was also used as the upper naturalness scale\footnote{In \cite{Barbieri:2006dq} the no-fine-tuning scale associated to the Higgs mass quadratic corrections was derived to be $\Lambda=1.3$~TeV and to be independent of the Higgs mass.}, and it was argued that with such a high scale one can no longer be certain that any new physics canceling the divergences will be observable at the LHC.

The plot includes all the IDMs from our scans that give a relic density in accordance with WMAP and pass all the constraints in Sec.~\ref{sec:2} except the Higgs bounds from LHC\footnote{Here we also impose the constraint in Eq. 17 of reference \cite{Barbieri:2006dq} even though we note that this does not qualitatively change the result.}. The solid blue line shows the results within the SM, and the mass range 115 to 129 GeV (the only span left for the SM Higgs given the current LHC limits) is marked as a thicker part of the solid blue line. This shows that $\Delta \approx 10$ for the SM. The kink on the blue curve around 350~GeV is when the fine-tuning goes from being dominated by $\Delta_{\Lambda_t}$ to being more sensitive to the Higgs cutoff $\Lambda_{h}$. We see that this measure $\Delta$ gives a large fraction of the IDMs (green circles) that are less fine-tuned than the SM ($\Delta \approx 10$), but also many models that are not.

\begin{figure}[t] 
\includegraphics[width=1 \columnwidth]{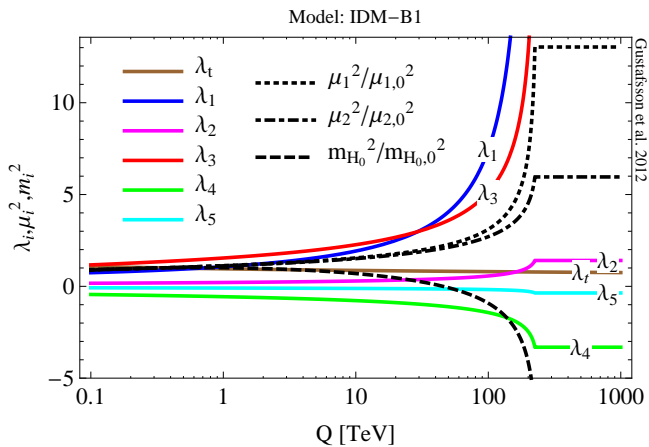}
\caption{The running of IDM parameters with energy scale $Q$ for one of our benchmark models (IDM-B1). Where the curves flatten out at around 150 TeV is when we terminate the calculation because the perturbativity limit of at least one $\lambda_i > 4\pi$ is reached. The $m^2_{H^0}$ and $\mu_i^2$ curves are normalized into units of their values $m^2_{H^0,0}$, $\mu_{i,0}^2$ at the scale $Q=m_h$.} 
\label{fig:running} 
\end{figure} 
\medskip
A similar measure to Eq.~\ref{eq:Delta_mu} was used in \cite{Casas:2006bd}, but with the running of the parameters up to the cutoff scale also taken into account. With $\Delta=\mbox{max}(|\Delta_{\Lambda_i^2}^{\mu_{1,2}}|)$ and the RG equations deduced from \cite{Haber:1993an,Branco:2011iw,Drozd:2012is}\footnote{All our renormalization conditions were set at $Q=m_h$.}, we find that with the fine-tuning condition $\Delta \leq5$ on $\mu_{1,2}^2$ our benchmark models are natural up to cutoff scales $\Lambda=1.0-2.4$ TeV. Figure \ref{fig:running} shows the running of the IDM parameters in the case of our benchmark model IDM-B1.
For comparison, the SM is now left natural up to $\Lambda=1.2$ TeV, with the SM Higgs mass is bound  to be $m_h<129$ GeV~\cite{Chatrchyan:2012tx}. This measure leaves half of our benchmark models less fine-tuned than the SM. In Figure~\ref{fig:tuning-scan} we also added the SM result, for $\Lambda=1.5$~TeV, when the RG running of couplings is included.
\begin{figure}[!t] 
\includegraphics[width=1 \columnwidth,height=0.695 \columnwidth]{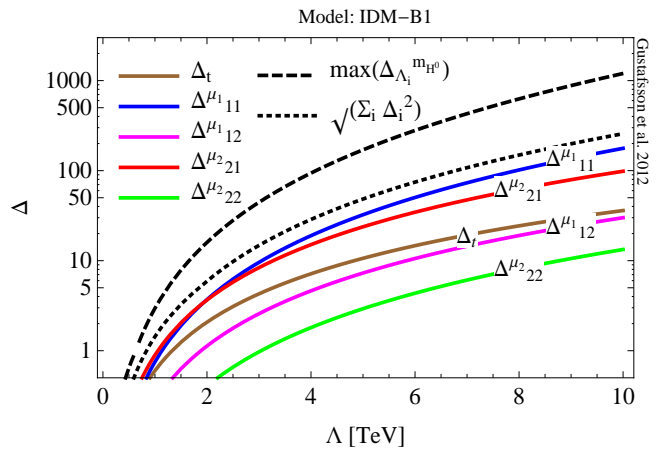}
 \caption{Fine-tuning measures, with RG effects included, for one of our benchmark models (IDM-B1). The notation for the plot legend is that $\Delta_{i}^X = \Delta_{\Lambda_i^2}^X$, and 
  $\sqrt{\sum_i \Delta_i^2} = \overline{\Delta}$ (as defined in the text).
 } 
 \label{fig:tuning} 
\end{figure} 

In \cite{Casas:2006bd} they used 
\mbox{$\overline{\Delta}
 = \text{max} \left(
\sqrt{
\sum_i
(\Delta_{\Lambda_i^2}^{\mu_{1,2}})^2+
(\Delta_{\lambda_i^2}^{\mu_{1,2}})^2
}\right)$}, 
where the contributions to $\Delta^{\mu_i}$ associated with $p_i=\lambda_i$ are also included. Here, the tuning with respect to
$\lambda_i$ has no significant impact (on our benchmark models), but we comment further on this below. In Figure~\ref{fig:tuning} we show this measure together with the individual contributions to the fine-tuning for our benchmark model IDM-B1.

We note, however, that in the case of our the benchmark models, the large quartic couplings are compensated by large negative values of $\mu_2^2$ to give small masses to the inert particles. This introduces an additional source of fine-tuning, even if each scale $\mu_1^2$ and $\mu_2^2$, associated with the two Higgs doublets, is individually not severely tuned. This we can incorporate by introducing a fine-tuning measure on, {\it e.g.}, the mass of the lightest inert particle
\bea
\label{eq:Delta_mH0}
\Delta_{\Lambda_i^2}^{m_{H^0}}
\!&\!\equiv\!&\!
\frac{\partial{\ln m_{H^0}^2}}{\partial{\ln{\Lambda_i^2}}}
\equiv
\!\frac{1}{m_{H^0}^2}\!
\frac{\partial{(\mu_2^2-\lambda_{H^0}\mu_1^2/2\lambda_1)}}{\partial{\ln{\Lambda_i^2}}}
\\\nonumber
\!&\!=\!&\!
   \frac{\mu_2^2}                      {m_{H^0}^2} \frac{\partial\ln{\mu_2^2}}{\partial{\ln{\Lambda_i^2}}}
 +\frac{\lambda_{H^0} v^2}{m_{H^0}^2} \frac{\partial\ln{\mu_1^2}}{\partial{\ln{\Lambda_i^2}}}
\eea
This reflects better the increased fine-tuning in models with high $\lambda_{H^0}$ and low $m_{H^0}$ values. It also cures the artificial large fine-tuning that arrises when $\mu_2^2$ goes through zero and drives $\Delta_{\Lambda_i^2}^{m_{H^0}}$ to infinity (any tuning around $\mu_2^2=0$ is irrelevant as it then does not contribute to any of the inert particle masses). The resulting \mbox{$\Delta=\mbox{max}(|\Delta_{\Lambda_i^2}^{m_{H^0}}|)$} are represented by crosses in Figure~\ref{fig:tuning-scan} (without RG improvement), as well as the {solid black} curve in Figure~\ref{fig:tuning} for IDM-B1 (including RG improvement). As our benchmark models come with rather large $\lambda_i$ this measure  typically leaves them less natural. $\Delta_{\Lambda}$ is less than 5
up to cutoff scales $\Lambda=0.4-1.4$ TeV when including the RG evolution. With this fine-tuning measure, our benchmark models can thus hardly be considered to be less fine-tuned than the SM.

We here also note that the sensitivity to variations in $p_i=\lambda_i$ could be significant already at tree-level. The tree-level contribution to
\bea
\Delta_{\lambda_i}^{m_{H^0}}\equiv
\frac{\partial{\ln m_{H^0}^2}}{\partial{\ln{\lambda_i}}}
\eea
already gives \mbox{$\Delta=\mbox{max}(|\Delta_{\lambda_i}^{m_{H^0}}|)\sim 6-25$} for our benchmark models, and is independent of $\Lambda_i$. This type of fine-tuning is however not directly related to the unknown contributions beyond the cutoff scale, and would be absent if we take our $\lambda_i$  to be fixed and known parameters for each model.

\section{Statistical measures}\label{appendix:B}
\begin{table*}
    \begin{center}
    \begin{tabular}{@{\extracolsep{\fill}} c||c|c|c||c|c|c||c}
    Model 	&	$3\sigma$, $\mathcal{G}$&  $3\sigma$ $P_\text{2}$=90\%, $\mathcal{G}$	&$3\sigma$ $P_\text{2}$=90\%, $\mathcal{P}$	& $5\sigma$, $\mathcal{G}$	& $5\sigma$ $P_\text{2}$=90\%, $\mathcal{G}$ & $5\sigma$ $P_\text{2}$=90\%, $\mathcal{P}$	& $P_1\!=\!P_2=95\%$, $\mathcal{P}$  	\\ \hline
    IDM-A1	&	720						& 1600					& 1700					& 2000				& 3400					& 3600						& 1000\\
    IDM-A2& 	240						&  590					& 630					& 680				& 1200					& 1300						& 380\\ \hline
    IDM-B1	& 	44						& 120					& 140					& 120				& 240					& 290						& 88\\
    IDM-B2	& 	44						& 120					& 140					& 120				& 240					& 290						& 88\\ \hline
    IDM-C1&	10\footnote{Requiring a signal expectation of at least 5 events gives the value quoted in parentheses in the first column.} (19)
    									& 36						& 44						& 28					& 66						&  90							& 30	\\				
    IDM-C2&	0.95\footnotemark[\value{mpfootnote}] (5.7) 
    									& 5.8						 & 8.3  		& 2.6\footnotemark[\value{mpfootnote}]	& 9.3					& 16							& 5.2\\
    IDM-C3	& 	11\footnotemark[\value{mpfootnote}] (19)	
    									& 38						& 49						& 30					&  70						& 96						& 30\\
    IDM-C4	& 	33						& 97						& 110					& 91					& 190					&  230  						& 70\\ 
      \end{tabular}
    \end{center}
\caption{Integrated luminosities $\mathcal L$, in units of fb$^{-1}$, required to detect our benchmark models under different statistical measures. The 3(5)$\sigma$ columns give the required integrated luminosity in order to observe a 3(5)$\sigma$ evidence (discovery) with a probability $P_2 = 90\%$, when the number of event counts is assumed to be Gaussian, $\mathcal G$, or Poisson, $\mathcal P$, distributed. In the columns with no $P_2$ value quoted, the commonly used criterion $S \geq 3(5)\sqrt{B}$ has been used ({\it i.e}\ the Gaussian approximation in Eq.~\ref{eq:gauss-w-margin}). In the last column we give the required luminosity to have a 95\% probability to exclude the models with at least 95\% confidence. (See the text for further information.)}
\label{tab:stat}
\end{table*}

It is desirable to have a statistical measure of the  integrated luminosity $\mathcal{L}$ expected to be 
needed to detect a signal with cross section $\sigma_S$ above a background with cross section $\sigma_B$. 

We denote the probability of observing $N$ or fewer events from a distribution with expectation value $X$ by
\beq
P_{\sss \rm D}(N;X), 
\eeq
where the index $\rm D$ distinguishes between different distributions. In the following,  $\rm D =G$ and $\rm P$ to denote Gaussian and Poisson statistics, respectively.
$B\equiv B(\mathcal L)=\sigma_{B} \mathcal L$ and  $S\equiv S(\mathcal
L)=\sigma_\text{S} \mathcal L$  denote the expectation values of the number of background and signal events, respectively.

\medskip
To claim that an observation of $N_{\text{obs}}$ events is an excess, {\it i.e.}\ to reject the null hypothesis of a background expectation $B$, it has to lie outside the interval specified by the background model's $P_1$ confidence level (CL). For a one-sided bound, this requires $N_\text{obs}\ge N_\text{min}(\mathcal{L})$, where $N_\text{min}$ is the minimum integer number satisfying
\beq
 P_{\sss \rm D}(N_\text{min};B)   \ge P_{1}.
\label{eq:n1}
\eeq 
For such a future observation to occur with a probability $P_{2}$, when the underlying true scenario has an expectation value $S+B$, it is required that $N_\text{min}$ also fulfills \mbox{$N_\text{min} \le N_\text{max}(\mathcal{L})$}, where $N_\text{max}$ is the maximum number satisfying
\beq
1-P_{\sss \rm D}(N_\text{max};S+B) \ge P_{2}.
\label{eq:n2}
\eeq
For a given distribution function $P_{\sss \rm D}(N;X)$, the system of equations \refeq{eq:n1}--\refeq{eq:n2} can then 
be solved to find the smallest required integrated luminosity $\mathcal{L}$ that has an integer solution $N$:
\beq
N_\text{min}(\mathcal{L},P_1) \le N \le N_\text{max}(\mathcal{L},P_2)
\label{eq:LN}
\eeq
Note that $P_D(N;X)$ are distribution functions, whereas $P_{1,2}$ are requested probabilities. 

It can be convenient to phrase the probabilities $P_{1,2}$ in terms of a corresponding number $n_{1,2}$ of standard deviations ($n$-$\sigma$) for a one-sided normal distribution. We define such a correspondence by
\beq 
P_{1,2} = \frac{1}{2} \left[ 1+\mathrm{erf} \left(  \frac{n_{1,2}}{\sqrt{2}} \right) \right], \label{eq:sigma}
\eeq
where erf is the Gaussian error function
\beq
\mathrm{erf}(x) = \frac{2}{\sqrt{\pi}} \int_0^x \mathrm{d}t\; e^{-t^2}.
\eeq
Equation~\ref{eq:sigma} thus defines what we refer to as an \mbox{$n$-$\sigma$} observation, independently of the type of distribution function $P_{\sss \rm D}$. For example 3(5)$\sigma$ corresponds to $1-P_{1,2} = 1.35\times10^{-3}(2.87\times10^{-7})$.

\bigskip
If the number $N$ of events is Poisson distributed, then the one-sided cumulative distribution function $P_{\sss\rm D}=P_{\sss\rm P}$ can be expressed as
\beq
P_{\sss \rm P}(N;X) = \frac{\gamma(N+1, X)}{\Gamma(N)}, \label{eq:poisson}
\eeq 
where $\Gamma$ and $\gamma$ are the ordinary and the lower incomplete gamma function respectively,  
\bea
\frac{\gamma(N+1,X)}{\Gamma(N)} 
 \;\operatornamewithlimits{=}^\text{for integer}_{N\ge0}\;&& \sum_{i=0}^N e^{-X}\frac{X^i}{i!}. 
 \label{eq:gamma}
\eea
Strictly speaking, $N$ can only take integer values -- as it represents the number of observed events -- and in general one can therefore not replace the inequalities with equalities in Eqs.~\refeq{eq:n1}--\refeq{eq:n2} and still find a solution. The analytical continuation ({\it i.e.}\ the gamma functions in Eq.~\refeq{eq:gamma}) can, however, be practical to have at hand, even though the final results should always derive from a solution with an integer $N$.

\smallskip
If the number $N$  of events is instead Gaussian distributed, then $P_{\sss\rm D}=P_{\sss\rm G}$ with 
\beq
 \label{eq:erf}
 P_{\sss \rm G}(N;X) = \frac{1}{2} \left[1 +  \mathrm{erf}\left(\frac{N-X}{\sqrt{2}\sigma}\right) \right],
\eeq 
where we take $\sigma=\sqrt{X}$ to coincide with a Poisson distribution for large $X$. In this case, Eq.~\refeq{eq:LN} can be written in a simple form. The $P_1$ ($n_1$-$\sigma$) CL one-sided upper limit on the background being smaller than the $P_{2}$  ($n_2$-$\sigma$) CL one-sided  lower limit on the signal plus background now reads
\bea
B + n_1 \sqrt{B} & \le N \le& S+B-n_2 \sqrt{S+B}.
\eea
The expression for the required signal $S$ can then be put into the following algebraic form (if we relax the requirement of $N$ being an integer):
\bea
\hspace{-0.5cm} S &\ge& n_1 \sqrt{B} +\frac{n_2}{2}\left[n_2 +\sqrt{4 B+4 \sqrt{B} n_1+n_2^2}\right]. \label{eq:Sineq}
\eea
Although $N$ should be an integer, we will follow common practice and leave out this additional requirement when we present results for Gaussian distributions in Table~\ref{tab:stat}. For $n_1=n$ and $n_2=0$, this gives the commonly used criterion for expecting an \mbox{$n$-sigma} detection 
\beq
S \ge n\sqrt{B}, \label{eq:gauss-w-margin}
\eeq 
which corresponds to a probability $P_2=50\%$ to observe the required $N_\text{obs}$ from a Gaussian distribution that, in fact, also spans over negative $N_\text{obs}$.
For \! $n_1\!=\!n_2\!=\!n$, Eq.~\refeq{eq:Sineq} gives the sometimes seen criterion \cite{Baer:1991yc}
\beq
 S \ge n^2 +2 n \sqrt{B}.
\eeq 

From these equations the minimum $\mathcal L$ is easily derived by substituting $B=\sigma_{B} \mathcal L$ and  $S=\sigma_\text{S} \mathcal L$.

\medskip
This defines our statistical measures to determine the expected integrated luminosity needed to observe a $n_1$-sigma detection with a probability $P_{2}$. 
Equivalently, this formalism also gives the expected integrated luminosity needed to exclude the signal expectation $S+B$ at the $P_{2}$ CL with a probability $P_{1}$.
 
\medskip
In Table~\ref{tab:stat} we present integrated luminosities required to detect our benchmark models with different probabilities $P_{1,2}$ under different assumed distribution functions $P_D$ for the number of event counts.


 

\begin{thebibliography}{00}
\bibitem{CMS:1TeV} 
The CMS collaboration also presented limits up to 1 TeV on SM like scalars \cite{CMS:1TeVb} that enables to fully exclude the remaining $> 600$ GeV mass region seen in Fig.~\ref{fig:AvailableRegion} for the SM Higgs boson in the IDM.
\end{thebibliography}
\end{document}